\documentclass[article]{aa}
\pdfoutput=1
\usepackage{epstopdf}
\usepackage{graphicx}
\usepackage[varg]{txfonts}
\usepackage{amsmath}
\usepackage{aas_macros}
\usepackage{float}
\usepackage{caption}
\usepackage{esdiff}
\usepackage{hyperref}

\bibpunct{(}{)}{;}{a}{}{,} % to follow the A&A style
\newcommand{\specialcell}[2][c]{%
  \begin{tabular}[#1]{@{}c@{}}#2\end{tabular}}

\makeatletter
\DeclareRobustCommand{\Oel}{O1\@ifnextchar,{}{\@ifnextchar.{}{ }}} %11
\DeclareRobustCommand{\Otwe}{O2\@ifnextchar,{}{\@ifnextchar.{}{ }}} %12
\DeclareRobustCommand{\Osi}{O3\@ifnextchar,{}{\@ifnextchar.{}{ }}} %6
\DeclareRobustCommand{\Ofo}{O4\@ifnextchar,{}{\@ifnextchar.{}{ }}} %4
\DeclareRobustCommand{\Oth}{O5\@ifnextchar,{}{\@ifnextchar.{}{ }}} %3
\DeclareRobustCommand{\Ofi}{O6\@ifnextchar,{}{\@ifnextchar.{}{ }}} %5
\DeclareRobustCommand{\Oon}{O7\@ifnextchar,{}{\@ifnextchar.{}{ }}} %1
\DeclareRobustCommand{\Ofot}{O8\@ifnextchar,{}{\@ifnextchar.{}{ }}} %14
\DeclareRobustCommand{\Ote}{O9\@ifnextchar,{}{\@ifnextchar.{}{ }}} %10
\DeclareRobustCommand{\Otht}{O10\@ifnextchar,{}{\@ifnextchar.{}{ }}} %13
\DeclareRobustCommand{\Otw}{O11\@ifnextchar,{}{\@ifnextchar.{}{ }}} %2
\DeclareRobustCommand{\Oni}{O12\@ifnextchar,{}{\@ifnextchar.{}{ }}} %9
\DeclareRobustCommand{\Ofit}{O13\@ifnextchar,{}{\@ifnextchar.{}{ }}} %15
\DeclareRobustCommand{\Ose}{O14\@ifnextchar,{}{\@ifnextchar.{}{ }}} %7
\DeclareRobustCommand{\Oei}{O15\@ifnextchar,{}{\@ifnextchar.{}{ }}} %8
\makeatother

\begin{document}
\title{The evolution of streams in a time-dependent potential}

\author{Hans J.T. Buist 
\and Amina Helmi}

\institute{Kapteyn Astronomical Institute, University of Groningen, P.O. Box 800, 9700 AV Groningen, The Netherlands\\ \email{buist@astro.rug.nl}}

\date{}
\abstract{We study the evolution of streams in a time-dependent
  spherical gravitational potential. Our goal is to establish what are
  the imprints of this time evolution on the properties of streams as
  well as their observability.  To this end, we have performed a suite
  of test-particle experiments for a host system that doubles its mass
  during the integration time and for a variety of initial conditions.
  In these experiments we found that the most striking imprint is a
  misalignment of $\sim 10^\circ$ in the angular location of the
  apocentres of the streams compared to the static case (and to the
  orbit of the centre of mass), which only becomes apparent for
  sufficiently long streams. We have also developed an analytic model
  using action-angle variables which allows us to explain this
  behaviour and to identify the most important signature of time
  evolution, namely a difference in the slope defined by the
  distribution of particles along a stream in frequency and in angle
  space. Although a difference in slope can arise when the present-day
  potential is not correctly modelled, this shortcoming can be by-passed 
  because in this case, streams are no longer straight lines in angle space, 
  but depict a wiggly appearance and an implausible energy gradient. The
  difference in slope due to time evolution is small, typically $\sim
  10^{-2}$ and its amplitude depends on the growth rate of the potential,
  but nonetheless we find that it could be observable if accurate full-space 
  information for nearby long streams is available. On the other hand,
  disregarding this effect may bias the determination of the present-day
  characteristics of the potential.}


\maketitle

\section{Introduction}

Throughout the history of the Milky Way and in the context of the
$\Lambda$ cold dark matter cosmogony, many dwarf galaxies must have been 
disrupted, leaving behind stellar streams \citep{Helmi1999}. Especially the Galactic halo probably contains many
such relics of this assembly history \citep{Helmi2008}.

Tidal streams were predicted for the first time in the seminal work of
\citet{ToomreToomre1972} who simulated the interactions of two galaxies. \citet{LyndenBell1995} and \citet{Johnston1996} put
forward the idea of tidal streams lurking in the Galactic halo, merely a few
years after the discovery of the Sagittarius dSph by
\citet{Ibata1994}. Streams and debris from the Sagittarius stream were found
several years later by \citet{Ivezic2000, Yanny2000, Ibata2001a}, and a full-sky view of the Sagittarius
stream was provided by \citet{Majewski2003}. Other examples are the globular cluster streams
\citep{Grillmair1995}, such as Palomar 5 \citep{Odenkirchen2001} and
NGC5466 \citep{Grillmair2006a}. A wealth of new streams have been found in
the Sloan Digital Sky Survey \citep[see e.g.][]{Grillmair2006b}, in the
`Field of Streams' \citep{Belokurov2006}. Streams have 
also been found with other galaxies \citep{MartinezDelgado2010,Martin2014}, most notably around M31 \citep{Ibata2001b}.

On the modelling side, much effort has been put into understanding the
dynamics of streams and using these to infer the Galactic potential.
Stream stars follow trajectories that are very similar to those of
their progenitors \citep{JinLyndenBell2007, Binney2008}, albeit
slightly offset. The evolution of streams is relatively simple in
action-angle coordinates \citep{Helmi1999, Tremaine1999}. Furthermore, 
streams appear as distinct clumps in integrals of motion space 
\citep{Helmi1998, HelmiDeZeeuw2000}, and they show the highest coherence in the true (underlying)
potential. This can be used to constrain its characteristic
parameters \citep{Penarrubia2012, Sanderson2014}. The
alignment of the angles and orbital frequencies can aid in determining
the potential, as proposed by \cite{Sanders2013b}.

The current cosmological model predicts significant evolution in the
mass content of galaxies and of their dark matter haloes through
cosmic time \citep{Springel2005}. This evolution may be directly
measurable using stellar streams, given their sensitivity to the
gravitational potential in which they are embedded
\citep{Johnston1999, Eyre2009, PriceWhelan2014, Bonaca2014}. A first study was presented by \citet{Penarrubia2006},
who concluded that there are no discernible effects of evolution on the
distribution of streams in the space of angular momenta and energy because
they only reflect the potential at the present day.  However,
\citet{Gomez2010a} found that the structure of streams in frequency
space does depict long-lasting signatures of time evolution. \citet{Bonaca2014} also
found several biases when they attempted to derive the characteristic parameters of a time
dependent potential using streams. They
attributed these biases partly to the time evolution of the potential in their simulations.

In the coming decade, the Gaia satellite \citep{Perryman2001},
successfully launched in 2013, will provide an unprecedented vast and detailed view of our Galaxy. With the correct understanding of streams, Gaia will allow us to address the assembly history of our Milky Way. This is the main
motivation of this paper. Our goal is to establish what kinds of observable imprints remain on  the stream properties and how to use them in recovering the evolutionary path of the Galactic potential. To this end we use the cosmologically motivated growth model of a spherical halo by \citet{BuistHelmi2014} and follow the evolution of streams both numerically and using the action-angle formalism.

This paper is organised as follows. In Sect.~\ref{sec:Methodology} we explain the simulations of streams in a time-dependent potential. We analyse the simulations in Sect.~\ref{sec:AnalysisSimulations} and develop an analytic model based on action-angles in the adiabatic regime. This allows us to identify the signature of time evolution on a stream in Sect.~\ref{sec:AnalyticModels}. In Sect.~\ref{sec:ObservationalProspects} we explore the prospects of observing this effect. We conclude in Sect.~\ref{sec:conclusions}.

\section{Methodology}
\label{sec:Methodology}

To understand the behaviour of streams in time-dependent potentials,
we take two complementary approaches. We first numerically study the
evolution of groups of test particles that initially resemble a satellite
galaxy. We then attempt to model this evolution using a formalism
based on action-angle variables. In Sect.~\ref{sec:sims-setup} we
describe the set-up of our numerical experiments, and in
Sect.~\ref{sec:aa} we briefly introduce the action-angle variables and
their properties.

\subsection{Simulation set-up}
\label{sec:sims-setup}
\subsubsection{Evolution of the gravitational potential}

\begin{figure}[!htbp]
\centering
\includegraphics[width=0.5\textwidth]{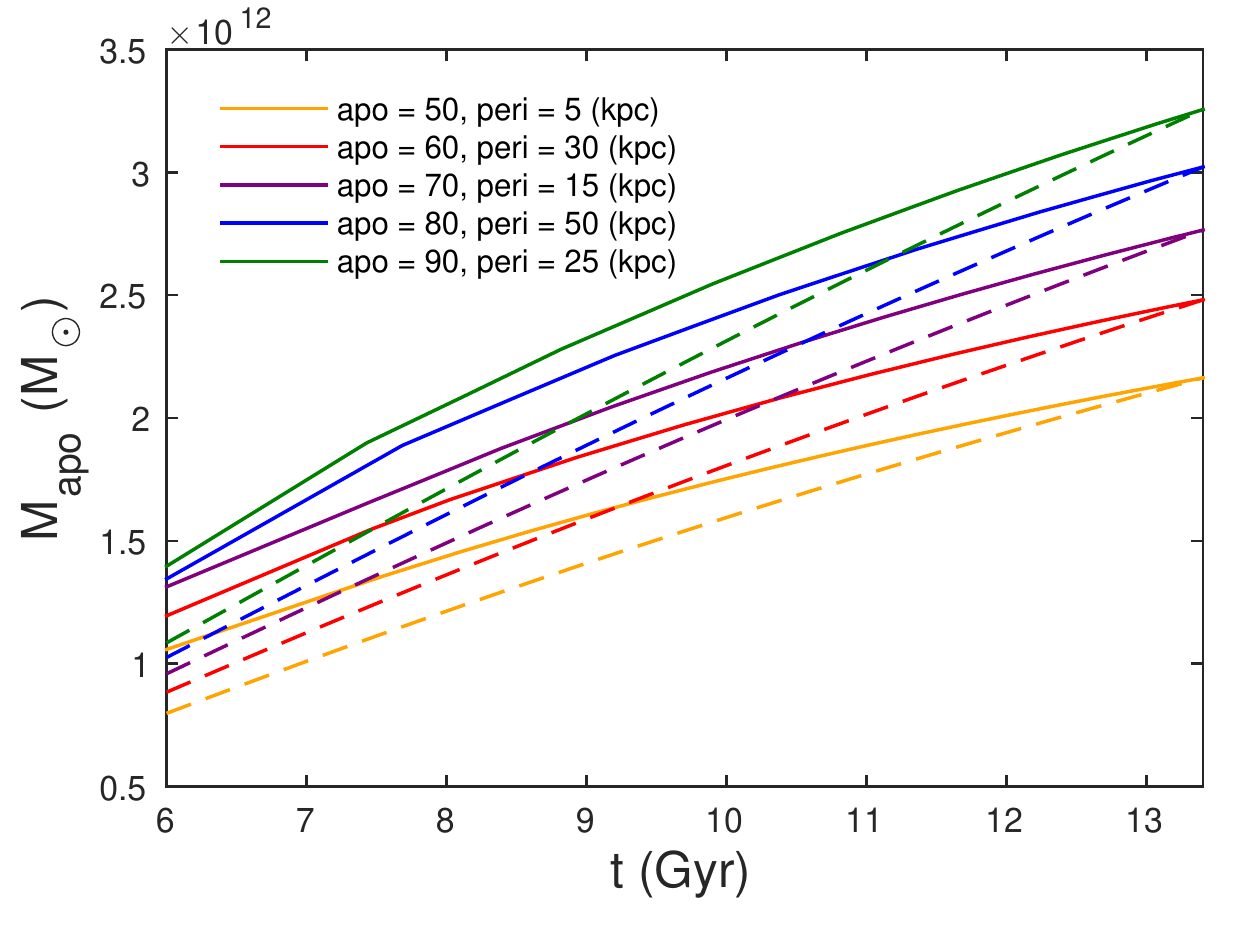}
\caption{\small Increase of the enclosed mass for several orbits for
  our time-dependent potential with a growth factor of $a_\textrm{g} = 0.8$. The
  dashed lines show the enclosed mass within the final value of their
  apocentre. Since the apocentric distances shrink with time, we show
  the mass enclosed within the time-dependent 
  apocentre values with solid lines. The integration time span for the streams is $t \sim 7-8 $ Gyr, and the enclosed mass approximately doubles over
  this timescale.}
\label{fig:massessshells}
\end{figure}
We modelled the evolution of a time-dependent
Navarro-Frenk-White potential \citep[][NFW]{NFW1997} using the
prescription models from \citet{BuistHelmi2014}. This model gives rise to an
inside-out smooth growth that well reproduces the average mass accretion history of
the Milky-Way-sized haloes from the Aquarius simulations
\citep{Springel2008, Wang2011}.  An important characteristic
of this model is that mass growth is positive at each radius (i.e.\ there is
no inward flow of mass between neighbouring mass shells). The scale 
radius $r_\textrm{s}$ and the scale mass
$M_\textrm{s}$ (mass enclosed within $r_\textrm{s}$) of such a halo vary according to 
\begin{align}
\begin{split}
  M_\textrm{s}(z) &= M_{\textrm{s},0} \exp{\left(-2 a_\textrm{g} z\right)}, \\
  r_\textrm{s}(t) &= r_{\textrm{s},0} \left(\frac{M_\textrm{s}(t)}{M_{\textrm{s},0}}\right)^{1/\gamma},
\end{split}
\end{align}
where $z$ is the redshift, and $M_{\textrm{s},0}$ and $r_{\textrm{s},0}$ the scale mass
and scale radius at the final time ($z=0$, in a cosmology with
$\Omega_\textrm{m}=0.29$ and $h_{100}=0.71$). We used $\gamma=2$ to grow
inside-out for an NFW potential \citep{BuistHelmi2014}. The
parameter $a_\textrm{g}$ determines the formation epoch, and higher values correspond to 
more recent formation. We here explore a range of values of $a_\textrm{g}$ up to $a_\textrm{g} = 0.8$, which would represent a quite recent
formation epoch. 

We modelled the evolution of streams in this time-dependent potential, but we also studied their behaviour in the static case. To make a fair comparison, we fixed  
the scale mass and scale radius at the final time to $M_\textrm{s}=5\times10^{11}$ $\textrm{M}_\odot$
 and $r_\textrm{s}=12$ kpc for all our experiments, 
independent of their time evolution. 

\subsubsection{Orbits}
\begin{table}[!htbp]
\small
\centering
\newcommand{\mc}[3]{\multicolumn{#1}{#2}{#3}}
\caption{\small Orbital properties at the final time}
\def\arraystretch{1.2}
\begin{tabular*}{0.5\textwidth}{rrrrrrr}
\hline
\hline
\specialcell[t]{Orbit\\} & \specialcell[t]{$r_\textrm{apo}$\\(kpc)} & \specialcell[t]{$r_\textrm{peri}$\\(kpc)} & \specialcell[t]{$L/L_\textrm{max}$\\} & \specialcell[t]{$\Omega_r$\\(rad/Gyr)} & \specialcell[t]{$\Omega_\phi$\\(rad/Gyr)} & \specialcell[t]{$\Omega_\phi/\Omega_r$\\} \\
\hline
% New ordering: 11 12 6 4 3 5 1 14 10 13 2 9 15 7 8
\Oel  & 90 & 5  & 0.22 & 11.38 & 6.96  & 0.61 \\ %11
\Otwe & 70 & 5  & 0.26 & 15.06 & 9.25  & 0.61 \\ %12
\Osi  & 60 & 5  & 0.29 & 17.79 & 10.96 & 0.62 \\ %6
\Ofo  & 50 & 5  & 0.33 & 21.54 & 13.32 & 0.62 \\ %4
\Oth  & 90 & 10 & 0.39 & 11.00 & 7.25  & 0.66 \\ %3
\Ofi  & 40 & 10 & 0.64 & 24.97 & 16.60 & 0.66 \\ %5
\Oon  & 70 & 15 & 0.61 & 13.69 & 9.45  & 0.69 \\ %1 
\Ofot & 30 & 20 & 0.96 & 25.89 & 18.15 & 0.70 \\ %14
\Ote  & 90 & 25 & 0.71 & 9.68  & 7.01  & 0.72 \\ %10
\Otht & 40 & 30 & 0.98 & 17.83 & 12.93 & 0.72 \\ %13
\Otw  & 60 & 30 & 0.90 & 13.21 & 9.68  & 0.73 \\ %2
\Oni  & 90 & 30 & 0.78 & 9.25  & 6.81  & 0.74 \\ %9
\Ofit & 50 & 40 & 0.99 & 13.32 & 9.89  & 0.74 \\ %15
\Ose  & 80 & 40 & 0.90 & 9.36  & 7.03  & 0.75 \\ %7
\Oei  & 80 & 50 & 0.95 & 8.53  & 6.49  & 0.76 \\ %8
\hline
\end{tabular*}
\label{tab:orbits}
\end{table}
{\begin{center}
\begin{table}[!htbp]
\small
\centering
\newcommand{\mc}[3]{\multicolumn{#1}{#2}{#3}}
\caption{\small Progenitor properties}
\def\arraystretch{1.2}
\begin{tabular}{@{\extracolsep{\fill}}lrr}
\hline
\hline
 & $\sigma_\textrm{pos}$ (pc) & $\sigma_\textrm{vel}$ (kpc/Gyr)\\
\hline
\textit{`Carina'} & $100$ & $5$\\
\hline
\textit{`Sculptor'} & $300$ & $10$\\
\hline
\textit{`Sagittarius'} & $700$ & $25$\\
\hline
\end{tabular}
\label{tab:progenitors}
\end{table}
\end{center}}

\noindent We placed the satellites on 15 different orbits as listed in Table
\ref{tab:orbits}. We chose this number to have a variety in distances,
orbital timescales (and hence the degree of adiabaticity), and eccentricity of the streams.
The apo- and pericentres at the final time were
chosen such that the particles spend most of their time in the outer
halo and beyond the scale radius of the potential. For a fair
comparison of each experiment the final 6D position of the satellite's 
centre of mass is the same for the static and time-dependent cases. 
We then integrated the orbits backwards in time for about 7-8
Gyr and the more circular orbits for 6-7 Gyr. This range was chosen to ensure that all orbits are bound at early times in the time-dependent case.

We chose to start our orbital integrations when the satellite's
centre of mass was at its first pericentre to represent the first
interaction with the Galaxy. This implies that the various orbits have
slightly different total integration times, but these only differ up to
a maximum of $0.5$ Gyr.  Within this integration time, the enclosed mass 
within a fixed radius increases on average with more than a factor 2 in our
halo, as seen in Fig.~\ref{fig:massessshells}. The actual orbits
experience a somewhat smaller increase in enclosed mass because
they respond to the halo growth by shrinking slowly.

\subsubsection{Progenitors}

We distributed the particles in the satellites assuming they follow an
isotropic Gaussian in position and in velocity space, characterised by
dispersions $\sigma_\textrm{pos}$ and $\sigma_\textrm{vel}$. We used 10,000 particles centred on the centre of mass of the satellite, which was
placed on the orbits described previously. We did not include
self-gravity in our simulations. We considered three different
progenitors, which we called `Carina', `Sculptor' and `Sagittarius' because they have properties reminiscent of these dwarf spheroidal satellites of the Milky Way \citep[see e.g.][see Table~\ref{tab:progenitors}]{Martin2008, Wolf2010}.

\subsection{Action-angle coordinates}
\label{sec:aa}

\subsubsection{Generalities}

A particularly useful description of the evolution of streams may be
obtained using action-angle coordinates \citep{Goldstein1950,
  Helmi1999, Tremaine1999}. In action-angle coordinates, the
actions are the momenta and the angles are the coordinates. 
The actions are integrals of
motion that are adiabatic, or in other words, invariant under slow
changes of the host potential.

In this section, we only work with a time-independent potential. 
For a single star, the actions can be found from
\begin{equation}
  J_i = \frac{1}{2\pi} \oint_{T_i} p_i dq_i 
\end{equation}
with $p_i$ the conjugate momentum of coordinate $q_i$, and $\oint_{T_i}$ to indicate that we integrate over a full period of coordinate $q_i$. In a spherically symmetric system the actions are the radial, latitudinal and azimuthal actions ($J_r$, $J_\theta$, and $J_\phi$ respectively), and the angles are $\theta_r$, $\theta_\theta$, and $\theta_\phi$, and represent the phase of the orbit in the $r$, $\theta$, and $\phi$ coordinates, respectively.

For only a few potentials algebraic expressions of the actions can be
derived, such as for the isochrone potential (of which the Kepler
potential and harmonic oscillator are limiting cases). Typically numerical methods are therefore required to find the actions
\citep{BinneyTremaine2008}. Furthermore, actions are only directly
available when the Hamilton-Jacobi equations are separable, as in the
case of spherical and
Staeckel potentials \citep{Goldstein1950, DeZeeuw1985}. In more general axisymmetric and triaxial potentials, approximations to the actions and angles must be used (see e.g.\ \citealt{KaasalainenBinney1994, McMillanBinney2011, Binney2012, Sanders2014a, Sanders2014b, Bovy2014}). 

\begin{figure}[!htbp]
\centering
\includegraphics[width=0.5\textwidth]{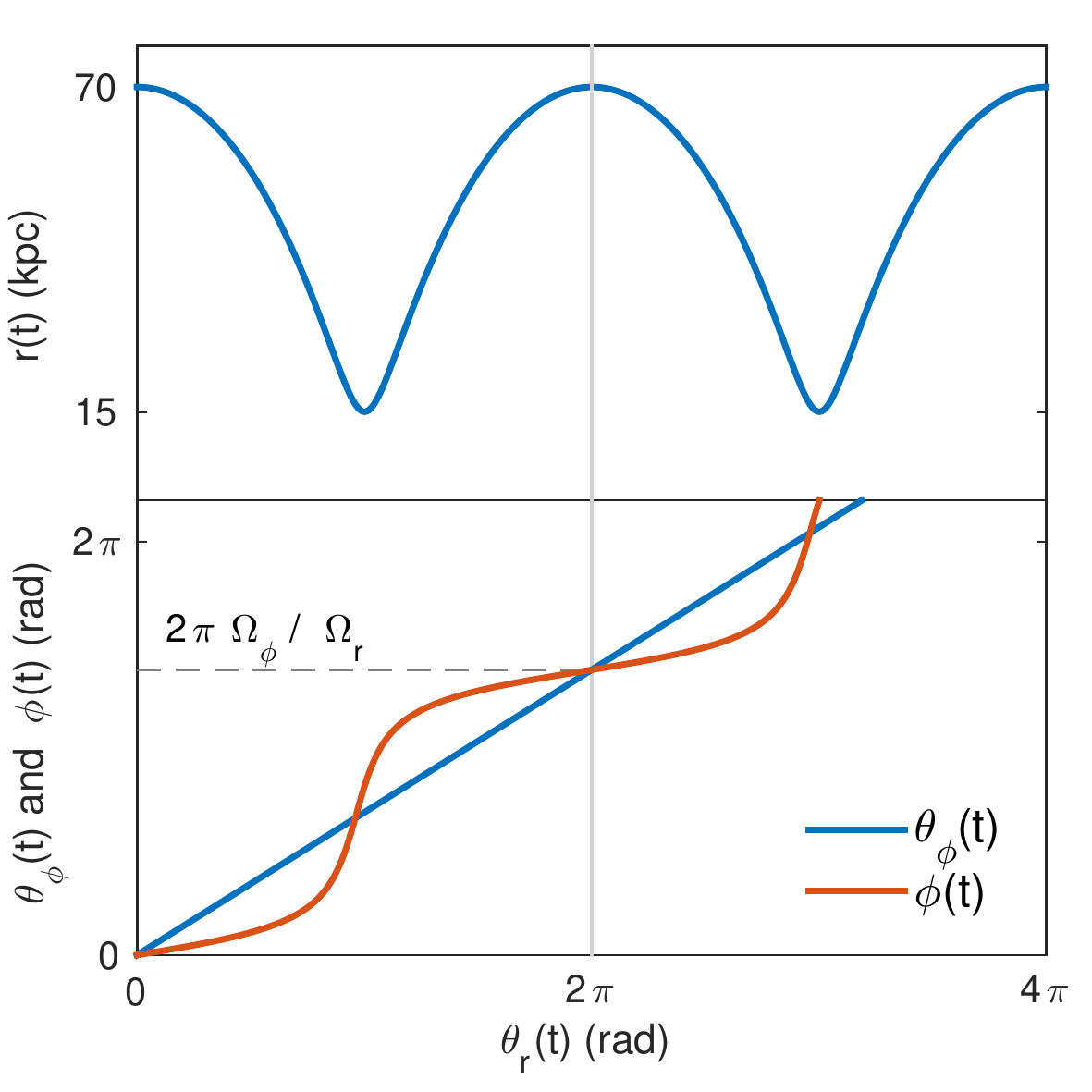}
\caption{\small Relation between the coordinates and the angles in the plane of the orbit. The top
  panel shows that the radial angle $\theta_r$ increases by $2\pi$ in
  one radial oscillation. In the bottom panel we have plotted $\phi$
  (red) and $\theta_\phi$ (blue) as functions of $\theta_r$. After one
  radial period, both angles reach the same value $2\pi \Omega_\phi / \Omega_r$.}
\label{fig:Thetas}
\end{figure}

Hamilton's equations for the $J_i$ and $\theta_i$ are
\begin{align}
\begin{split}
  \dot{J}_i      &= \frac{\partial H(\vec{J})}{\partial \theta_i} = 0, \\
  \dot{\theta}_i &= \frac{\partial H(\vec{J})}{\partial J_i} \equiv \Omega_i(\vec{J}),
\end{split}
\label{eq:eqnmotion}
\end{align}
and $H=H(\vec{J})$ only because the actions are invariant. The $\Omega_i$ are constants that correspond to the orbital frequencies and also only depend on the $J_i$. The resulting equations of motion for the $\theta_i$ are 
\begin{equation}
  \theta_i = \Omega_i t + \theta_i(0),
\end{equation}
with $\theta_i(0)$ the initial phase of $i$-th coordinate. 

To determine the angles, we used a canonical transformation of the
second kind $W(\vec{q},\vec{J})$ \citep{Goldstein1950,
  BinneyTremaine2008}, such that
\begin{equation}
  p_i = \frac{\partial W(\vec{q},\vec{J})}{\partial q_i} ; \ \ \ \ \ \ \theta_i = \frac{\partial W(\vec{q},\vec{J})}{\partial J_i}.
\label{eq:coordinatetransformation}
\end{equation}
The generating function for the transformation in a spherical potential is given by (adapted from \citealt{BinneyTremaine2008})
\begin{equation}
\begin{split}
  W(\vec{q},\vec{J}) &= W_\phi(\phi, \vec{J}) + W_\vartheta(\vartheta, \vec{J}) + W_r(r, \vec{J}) \ \\
                           &= \int_{\phi_\text{min}}^{\phi} d\phi p_\phi(J_\phi) + \int_{\vartheta_\text{min}}^\vartheta d\vartheta\, p_\vartheta(J_\phi, J_\vartheta) + \int_{r_\textrm{peri}}^r dr\, p_r(J_r, J_\phi, J_\vartheta),
\end{split}
\label{eq:generatingfunction}
\end{equation}
where the integration is along the trajectory of a particle in phase space. We note that the three parts of the generating function are indefinite versions of the action integrals without a factor of $2\pi$, and each increases therefore by $2\pi J_i$ in their corresponding periods $2\pi/\Omega_i$. This equation has an algebraic expression in the case of an isochrone potential and has to be solved numerically otherwise. For the full integrals needed to compute $\theta_i$ in a spherical potential we refer to Appendix \ref{sec:AppendixA}.

\begin{figure*}[!htbp]
\centering
\noindent\makebox[\textwidth]{
\includegraphics[width=1.0\textwidth]{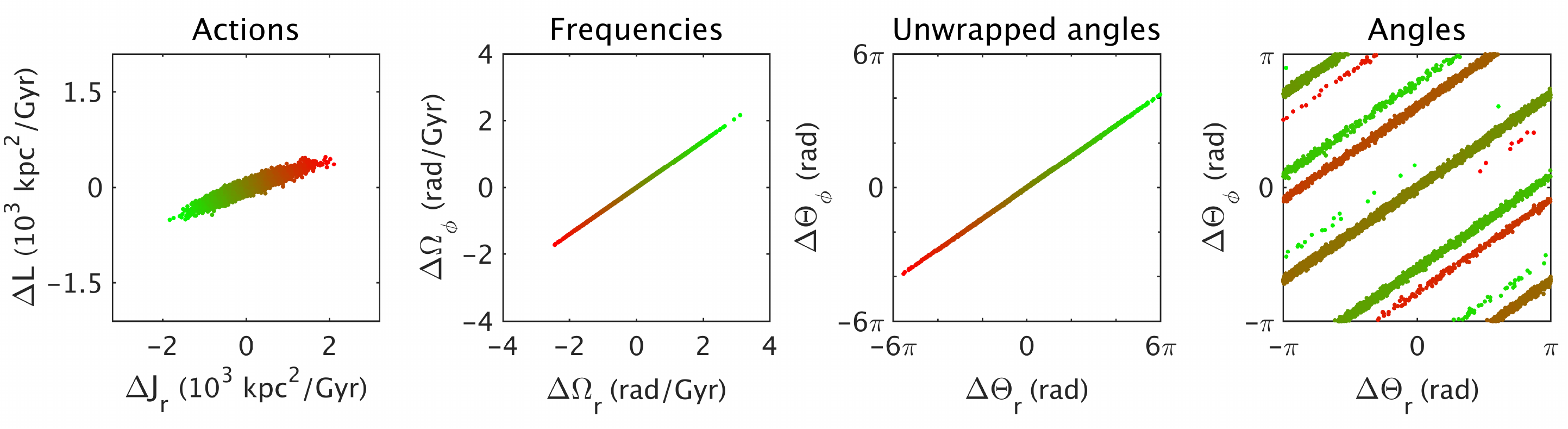}}
\caption{\small Distribution in action (left), frequency (second panel) and angle space (right) for particles in a stream, and centred on the centre of mass of the progenitor system. The colours represent the energy gradient, where red is for the trailing and green for the leading arms.}
\label{fig:Delta3}
\end{figure*}

In Fig.~\ref{fig:Thetas} we show an example of how the radial and azimuthal angles ($\theta_r$
and $\theta_\phi$) are related to the radial and
azimuthal coordinates ($r$ and $\phi$) in the orbital plane. Since $W_\phi =
\phi J_\phi$ and using Eqs.~(\ref{eq:coordinatetransformation}) and
(\ref{eq:generatingfunction})\footnote{$\displaystyle \frac{\partial W_\theta}{\partial J_\phi} = 0$ on the orbital plane.}
\begin{align}
  \theta_\phi &= \frac{\partial W}{\partial J_\phi} = \phi + \frac{\partial W_r}{\partial J_\phi} .
\end{align}
The term $\displaystyle \frac{\partial W_r}{\partial J_\phi}$ has a non-secular oscillation with frequency $\Omega_r$ and vanishes at apo- and pericentre ($\theta_r=\left\{0, \pm \pi \right\}$). The secular behaviour of $\phi$ and $\theta_\phi$ with time is  
therefore the same, as can also be seen in Fig.~\ref{fig:Thetas}.

\subsubsection{Streams in action-angle coordinates: time-independent case}

A stream is created when particles drift away from the progenitor, which means we should look at the relative phase with respect to the progenitor centre of mass. For example, the relative azimuthal phase of the $k$-th particle is
\begin{equation}
\begin{split}
\theta_\phi^k(t) - \theta_\phi^\textrm{cm}(t)   =  \Delta \theta_\phi^k(t) &= \Delta \theta_\phi^k(0) + \Delta\Omega_\phi^k t \\
  &\approx \Delta\Omega_\phi^k t ,
\end{split}
\label{eq:anglespreads}
\end{equation}
where $\Delta \theta_\phi^k(0)$ is the separation in angles at the
initial time, and where the second line corresponds to the case
in which the progenitor is small and the stream has evolved for a sufficiently long time. 

The $\Delta\Omega_\phi^k = \Omega_\phi^k - \Omega_\phi^\textrm{cm}$ depends on
the distribution in action space $J_i^k$, such that a broad action
spread results in a wide range of frequencies. This implies that
larger progenitors generate longer streams than smaller ones if
integrated for the same amount of time.

In Fig.~\ref{fig:Delta3} we show the structure of a stream in
action-angle coordinates for the time-independent case.  The colours
represent the energy gradient, such that green represents the leading
tail and red the trailing tail.  The left panel shows the
action distribution, which depends both on the initial conditions $\sigma_\textrm{pos}$ and $\sigma_\textrm{vel}$ and on 
the initial orbit\footnote{Although here we plot $L = J_\theta + |J_\phi|$, but in the orbital plane $J_\theta = 0$.}. This distribution also remains invariant in time in an adiabatically evolving potential. The second-left panel of shows the structure in frequency space, while the third corresponds to angle space.  Since angles are $2\pi$ periodic variables, in the right most panel we show the angles modulo $2\pi$. The different streaks signify the number of radial and azimuthal wraps the stream has. We note that the slope of the
lines in frequency and in angle space are the same in this time-independent
potential \citep{Sanders2013b} when the initial spread in angles can be neglected\footnote{This equation holds for every particle that is released at the same time from the progenitor.}:
\begin{equation}
\label{eq:sameSlope}
\frac{\Delta \theta_\phi}{\Delta \theta_r} = \frac{\Delta\Omega_\phi}{\Delta\Omega_r} = cst .
\end{equation}
This directly shows why all particles are
distributed along straight lines with the same slope in angle and in frequency space.

\section{Analysis of the test-particle simulations}
\label{sec:AnalysisSimulations}
\begin{figure*}[!htbp]
\centering
\vspace*{-1cm}
\noindent\makebox[\textwidth]{
\includegraphics[width=1.05\textwidth]{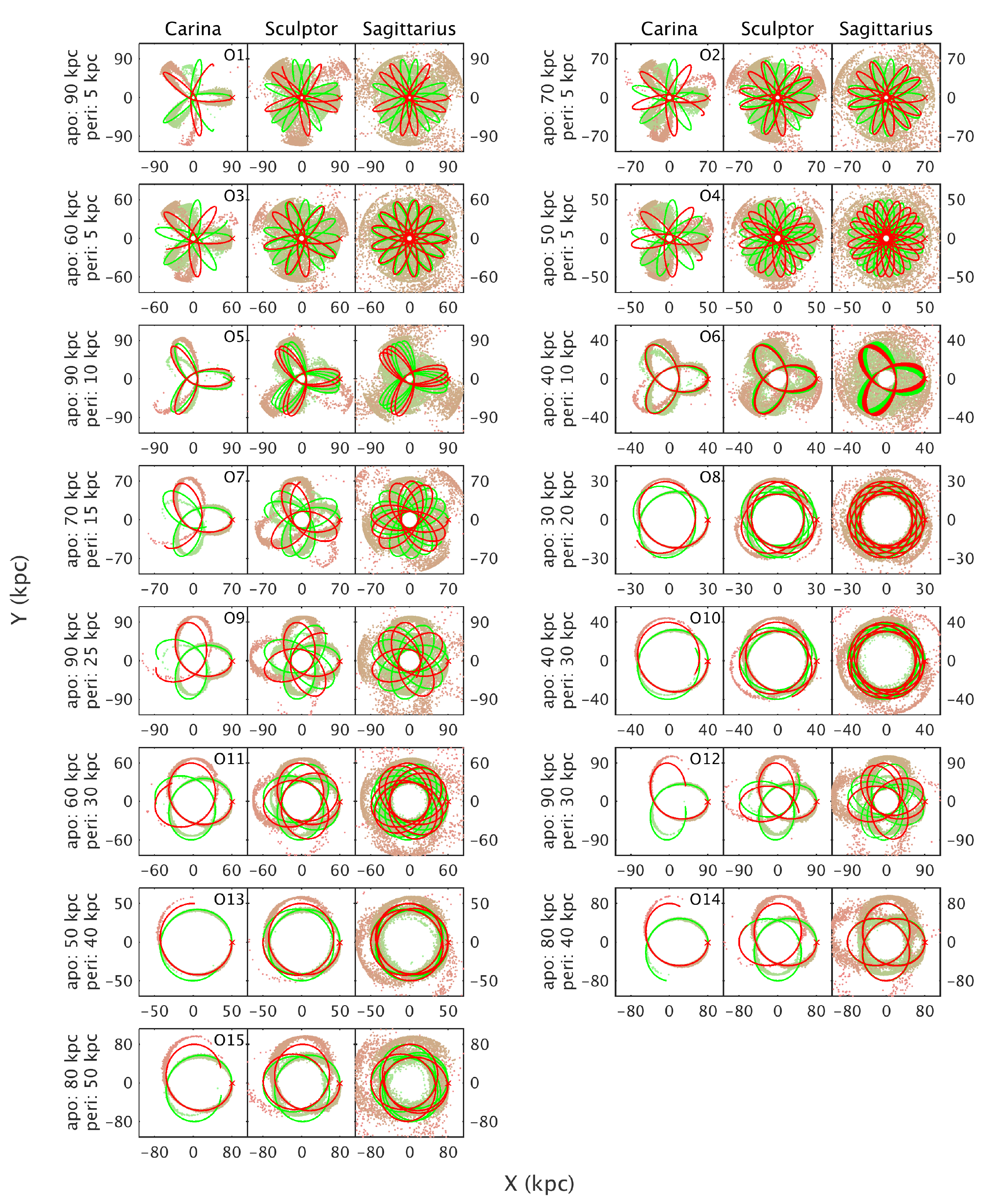}}
\caption{\small Streams from three progenitor sizes for each of the 15 orbits
  evolved in the time-dependent $a_\textrm{g}=0.8$ potential. The left, middle,
  and right panels correspond to the `Carina', {`Sculptor', and 
  `Sagittarius'}-like progenitors, respectively (see Table
  \ref{tab:progenitors} for their properties).  Particles in green are more bound to the host galaxy
  (leading arm), while those in red are less bound (trailing arm) than the
  progenitor. The dashed curves are the progenitor orbits
  evolved in today's potential, with the green integrated forward in time
  for the leading arm and red backwards for the trailing arm. The current position of 
  the progenitor is indicated with a red cross.}
\label{fig:XY3}
\end{figure*}

\begin{figure*}[!htbp]
\centering
\noindent\makebox[\textwidth]{
\includegraphics[width=0.9\textwidth]{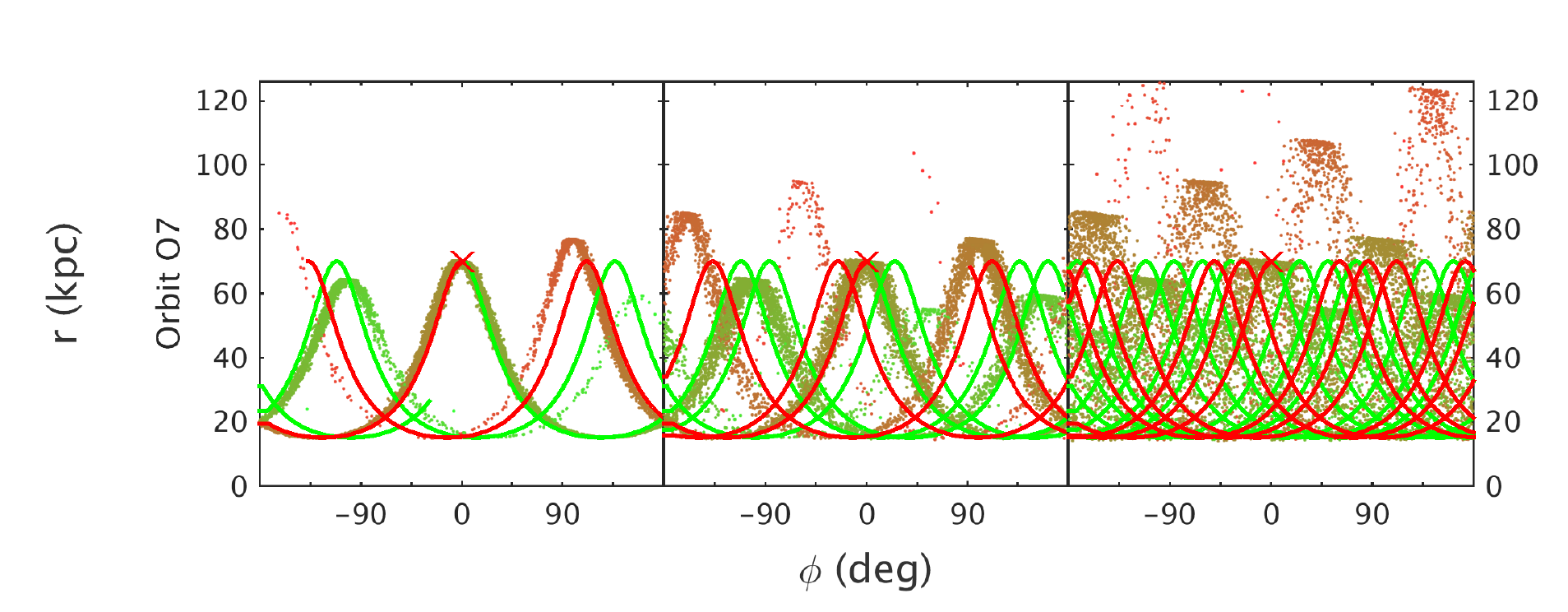}}
\caption{\small Radial vs angular distribution of particles in the
  orbital plane for experiment \Oon for the different progenitors in the
  time-dependent case. The dashed curves show the orbit of the centre
  of mass of the system integrated long enough to roughly reproduce
  the lengths of the streams. The variations in the apocentric distances are
  a reflection both of time evolution and of the energy gradient
  present along a stream. The difference in angular location of the
  apocentre, on the other hand, is a clear imprint of time evolution, as can be seen by comparing to e.g.\ Fig.~\ref{fig:rphi}.}
\label{fig:rphi1}
\end{figure*}

\begin{figure*}[!htbp]
\centering
\vspace*{-1cm}
\noindent\makebox[\textwidth]{
\includegraphics[width=1.05\textwidth]{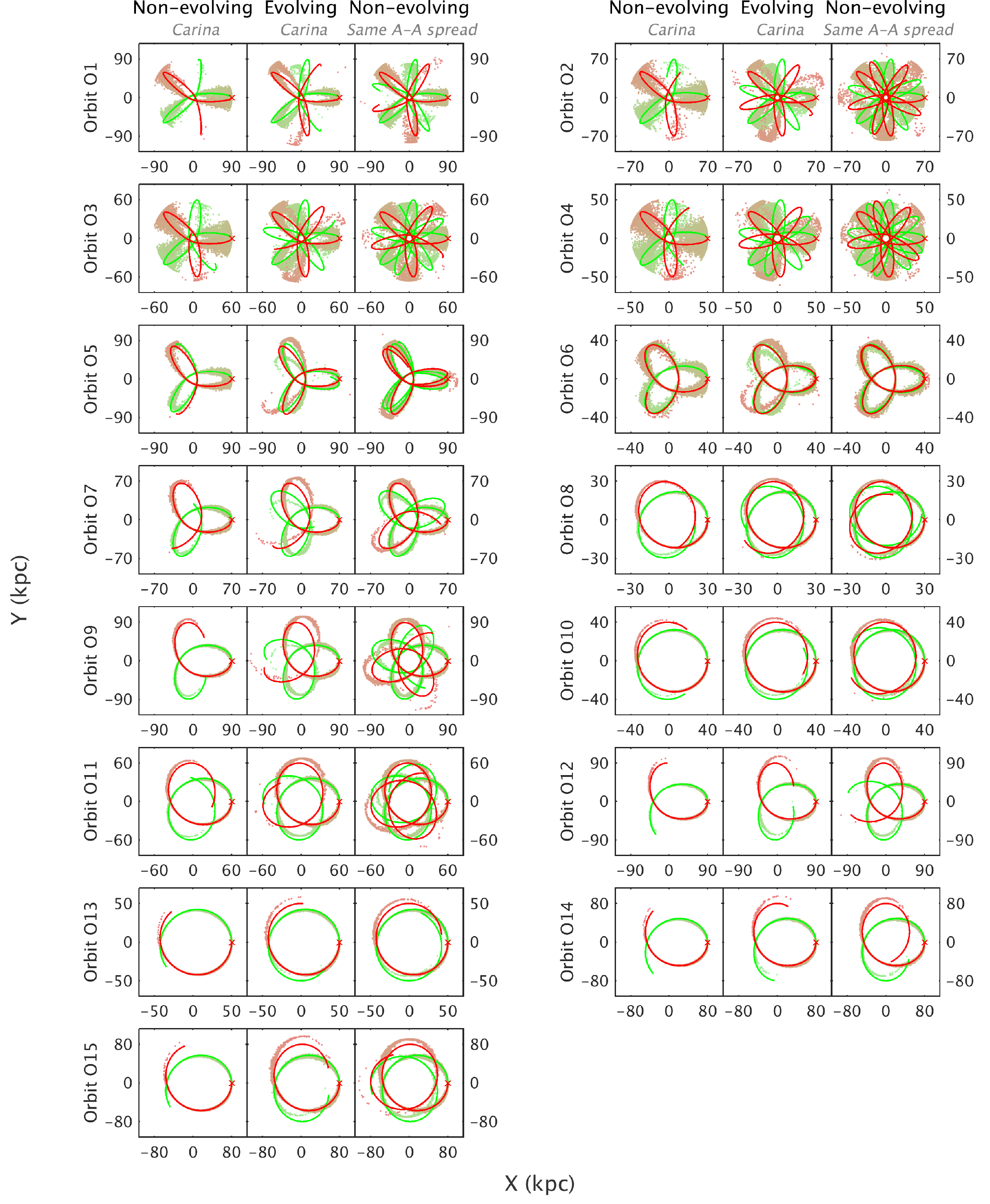}}
\caption{\small Streams from the Carina progenitor in the evolving and in the non-evolving potential. The streams in the left and centre have the same initial distribution in configuration and velocity space. Those in the right panels have the same initial action-angle distribution as the time-evolving potential (middle panel), but are evolved in a static potential. The colour coding is the same as in the previous figure. The stream-orbit misalignment can be seen by looking at the azimuthal angle of the petals starting from one radial period away from the progenitor, whose position is indicated with a red cross.}
\label{fig:xyexample}
\end{figure*}

\begin{figure*}[!htbp]
\centering
\noindent\makebox[\textwidth]{
\includegraphics[width=0.9\textwidth]{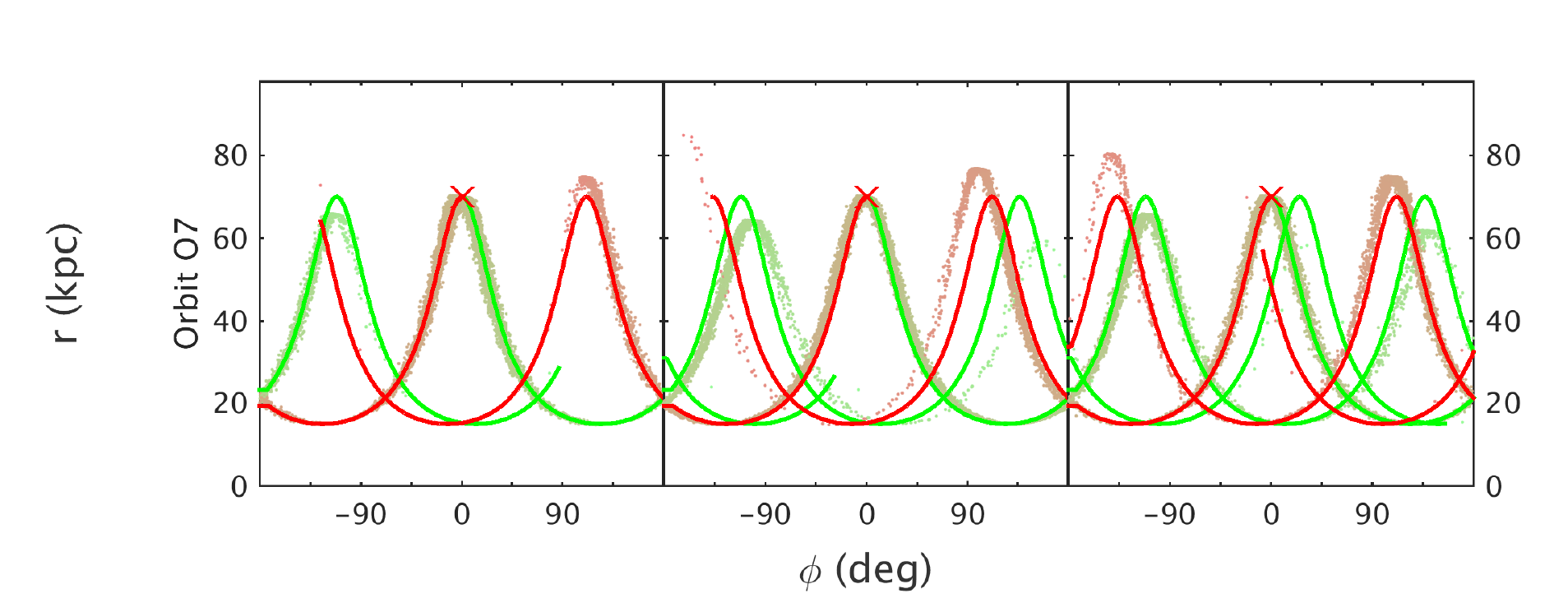}}
\caption{\small Radial vs angular distribution of particles in the
  orbital plane for experiment \Oon for the Carina-like progenitor in
  the static case (left), time-dependent case (middle), and static case
  with the same initial action-angle distribution as in the middle
  panel. The dashed curves show the orbit of the centre of mass of the
  system integrated in the present-day potential for the left and
  right panels, and in the evolving potential for the middle
  panel. For the static cases, the variations in the apocentric
  distances are a reflection of the energy gradient present along the
  stream, while a second effect is present in the middle panel because
  of time evolution. In this case, the difference in angular
  location of the apocentre is also much more pronounced and systematically
  increases for older wraps.}
\label{fig:rphi}
\end{figure*}

\begin{figure*}[!htbp]
\centering
\noindent\makebox[\textwidth]{
\includegraphics[width=1.05\textwidth]{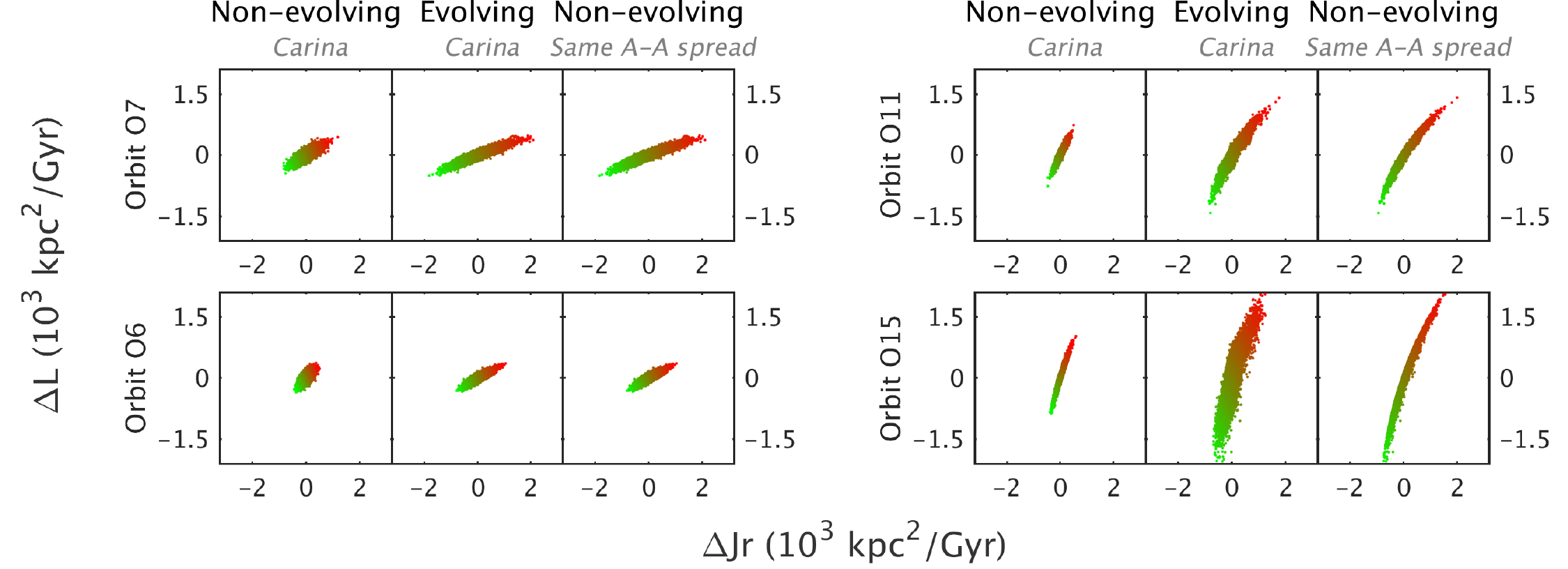}}
\caption{\small Action-space distributions for a sample of four streams from Fig.~\ref{fig:xyexample}, measured with respect to the centre of mass of the progenitor at the final time. The colours of the particles represent the energy gradient, such that green is the leading arm and red the trailing arm. In the static cases, the initial and final distributions coincide exactly, but in the time-dependent case there can be small deviations if the orbit is not in the adiabatic regime, as can be seen by comparing the middle and right panels for orbits \Otw and \Oei. The left panels show a smaller spread because the time-dependent cases start farther outwards and in an initially much lighter potential.}
\label{fig:JrL}
\end{figure*}

The results of the simulations for the three progenitors on the
various orbits for the time-dependent potential with growth factor
$a_\textrm{g} = 0.8$ are shown in Fig.~\ref{fig:XY3}. The colour coding
indicates the leading (green) and trailing (red) arms of the
stream. For comparison we have over-plotted the orbit of the
progenitor integrated forward and backwards in time in the present-day
potential.  The time of integration for that orbit is between 1 to 3
Gyr and is chosen to roughly match the length of the corresponding
stream. Note that this is much shorter than the time of integration of
the particles that form the stream, which is approximately 8 Gyr.

Figure \ref{fig:XY3} clearly shows that larger progenitors give rise
to longer and wider streams.  The smaller (Carina) progenitor
generates thin, short streams, which makes the separate loops of the
stream easily recognizable by eye, while the streams from the larger
progenitors, depending on the specific orbit considered, are wider and
longer, much more phase mixed, and sometimes difficult to discern. As we show below, a clear imprint of the time evolution of the potential is apparent in
the angular location and in the variation of the maximum distance of
each `petal' of a stream (the latter being related to both the
shrinking of the progenitor's orbit because of mass growth and
to the energy gradient along a stream). This effect is more easily
discernible for heavier progenitors since they produce more loops,
that is, longer streams, as can be seen in Fig.~\ref{fig:rphi1}. On the
other hand, thinner streams have the advantage that they depict their
mean orbit much more clearly, which is the reason we mainly focus on the
`Carina' progenitor in what follows.

The differences in the evolution for time-dependent and static
potentials are shown in Figs.~\ref{fig:xyexample} and \ref{fig:rphi} for the Carina-like
progenitor, depicted as before on the orbital plane of the
progenitor. We note that in all cases the stream in the time-dependent
potential is longer than in the static case, as can be seen by 
comparing the left and middle panels. 

The length of a stream depends both on the orbital period and
on the initial extent in phase- and configuration space of the progenitor. In the
time-dependent potential the enclosed mass at initial times is lower, and
the progenitor starts farther out. This results in lower orbital frequencies, or longer periods. Since the rate of divergence of particles in a stream in a
spherical potential is proportional to $(t/P)^2 \propto (\Omega t)^2$
\citep{Helmi1999}, this would imply that streams should actually be
shorter for a fixed integration time $t$. However, in a time-dependent potential the spread in actions (or integrals of motion such as the energy) is broader for given initial $\sigma_\textrm{pos}$ and $\sigma_\textrm{vel}$ (see Fig.~\ref{fig:JrL}).

For an initially `compact' ensemble of particles, we may use a linear transformation between action-angle coordinates and Cartesian coordinates to derive the initial spread in action-angle space:
\begin{equation}\renewcommand\arraystretch{1.2}
\left[\begin{matrix} \Delta \boldsymbol\theta \\ \Delta \vec{J}
\end{matrix}\right] = \left[ \renewcommand\arraystretch{2.2}\begin{matrix}  \dfrac{\partial\boldsymbol\theta}{\partial\vec{q}} & \dfrac{\partial\boldsymbol\theta}{\partial\vec{p}} \\ \dfrac{\partial\vec{J}}{\partial\vec{q}} & \dfrac{\partial\vec{J}}{\partial\vec{p}} \end{matrix} \right] \cdot \left[ \renewcommand\arraystretch{1.2} \begin{matrix} \Delta \vec{q} \\ \Delta \vec{p}
\end{matrix}\right] \equiv \vec{T} \cdot \left[ \renewcommand\arraystretch{1.2} \begin{matrix} \Delta \vec{q} \\ \Delta \vec{p}
\end{matrix}\right] ,
\label{eq:lintransformation}
\end{equation}
where the elements of the transformation matrix $\vec{T}$ will
depend on second derivatives of the generating function evaluated at
the centre of mass of the progenitor \citep{Helmi1999}, as described in
detail in Appendix \ref{sec:AppendixB}. We see that many of the terms
depend on $1/\Omega_r$, which is larger in a shallower potential and
farther out in the potential, confirming that the initial spread in actions is greater. Therefore the particles in the progenitor integrated
in the evolving potential can spread more and give rise to longer and
wider streams.

In our initial set-up we took the progenitors to have the same
$\sigma_\textrm{pos}$ and $\sigma_\textrm{vel}$. However, their initial distribution in
action-angle space is different for the static and time-dependent
potentials, as discussed above. This implies that their final
distribution in action (or energy) space will also be different. This is
shown for the progenitor placed on two of our orbits in
Fig.~\ref{fig:JrL}. Here we have plotted the distribution at the final time of $\Delta J_r$ and $\Delta L$ in the top panels for the static (left) and time-dependent
(middle) potentials for the `Carina'-like progenitor.

An alternative is to consider that the particles have
the same initial $\Delta J_r$ and $\Delta L$, whether they are evolved
in a static or time-dependent gravitational potential. To establish the effect of
such a change in the initial configuration, we set up progenitors with
these properties and evolved them in the static potential. The right panels 
of Fig.~\ref{fig:JrL} show the distribution of particles evolved in the static potential, 
but set up with the same initial conditions in action-angle space as in 
the time-dependent case. If all orbits were in the adiabatic regime, the middle 
and right panels should give identical results. However, some of the orbits are 
not, which can be seen in the streams from orbits \Otw and \Oei, where the action 
distributions are not exactly the same.

The streams resulting from these different initial conditions 
evolved in the time-independent potential are shown in the right
panels of Fig.~\ref{fig:xyexample} for each of the orbits. As
anticipated, the streams in the time-dependent potential (middle
panels) are now shorter than the streams in the time-independent
potential starting from the new initial conditions. This
characteristic is now solely due to the evolution of the potential.

An interesting feature visible in Figs.~\ref{fig:xyexample} and \ref{fig:rphi} that was 
mentioned earlier is that the stream-orbit misalignment differs in
the static and evolving cases. For example, as can be seen for 
experiment \Oon, the mean orbit traces the stream in the
time-independent case relatively well. On the other hand, it is clear that the angular position of the rosette petal of the stream and the progenitor orbit
at one radial period behind (or ahead) are offset from each other. This
is apparent in all experiments, to a lesser or greater degree. This
offset is systematic, can be as large as 10 degrees, and is a first
direct indication of the effect of time evolution on the gravitational
potential.  

\section{Analytic models}
\label{sec:AnalyticModels}
\subsection{Action-angle coordinates in an adiabatically evolving potential}

In this section our aim is to extend the use of action-angles to
an adiabatically changing spherical potential. The action-angles are still a valid
canonical coordinate system in a time-dependent system, but the
equations of motion in action-angle coordinates are more
complicated, as we show below.

In the models we have considered so far, the gravitational potential
is made time dependent by making its characteristic parameters, such
as mass and scale, a function of time. Its overall shape or functional
form remain the same. Therefore angular momentum is still conserved
because of the assumed spherical symmetry, but the radial action $J_r$
may vary with time.

For a time-dependent potential, the generating function that allows
the transformation between Cartesian and action-angle spaces is an
explicit function of time
\begin{equation}
  W(\vec{q},\vec{J}, t) = W(\vec{q},\vec{J}, \boldsymbol\alpha(t)) ,
\end{equation}
where the parameters of the potential are in the vector
$\boldsymbol\alpha$. The appropriate Hamiltonian in action-angle
coordinates then becomes
\begin{equation}
  H'(\boldsymbol\theta,\vec{J}) = H(\vec{J},\boldsymbol\alpha) + \frac{\partial W}{\partial t}(\vec{q},\vec{J}, \boldsymbol\alpha) = H(\vec{J},\boldsymbol\alpha) + \dot{\boldsymbol\alpha} \frac{\partial W}{\partial \boldsymbol\alpha}(\vec{q},\vec{J}, \boldsymbol\alpha),
\label{eq:higherorderterms}
\end{equation}
which is the original Hamiltonian perturbed by the partial time
derivative of the generating function \citep[section
11-7]{Goldstein1950}. This extra term naturally vanishes in the
time-independent case. The new equations of motion are
\begin{align}
\begin{split}
  \dot{J}_i &= -\frac{\partial H'}{\partial \theta_i} = -\dot{\boldsymbol\alpha} \frac{\partial}{\partial \theta_i} \frac{\partial W}{\partial \boldsymbol\alpha}(\vec{q},\vec{J}, \boldsymbol\alpha), \\
  \dot{\theta}_i &= \frac{\partial H'}{\partial J_i} = \Omega_i(\vec{J},\boldsymbol\alpha) + \dot{\boldsymbol\alpha} \frac{\partial}{\partial J_i} \frac{\partial W}{\partial \boldsymbol\alpha}(\vec{q},\vec{J}, \boldsymbol\alpha).
  \label{eq:neweqnmotion}
\end{split}
\end{align}
Compared to the original equations of motion given in
Eq.~(\ref{eq:eqnmotion}), there are a few correction terms that depend
on the rate of change of the characteristic parameters of the
potential $\dot{\boldsymbol\alpha}$. 

Since in general we are interested in the mean
increase in the actions and angles, we take an average
over one period, where we assume that $\boldsymbol\alpha$ changes
little over one period \citep{Goldstein1950, Vandervoort1961}
\begin{align}
\begin{split}
   \left< \dot{J}_i \right> &= \frac{1}{T_i} \int_{T_i} -\dot{\boldsymbol\alpha} \frac{\partial}{\partial \theta_i} \frac{\partial W}{\partial \boldsymbol\alpha}(\vec{q},\vec{J}, \boldsymbol\alpha) dt \\ &\approx -\frac{\dot{\boldsymbol\alpha}}{T_i} \int_{T_i} \frac{\partial}{\partial \theta_i} \frac{\partial W_i}{\partial \boldsymbol\alpha}(\vec{q},\vec{J}, \boldsymbol\alpha) dt +\mathcal{O}(\dot{\boldsymbol\alpha}^2,\ddot{\boldsymbol\alpha}).
\end{split}
\label{eq:averageJ}
\end{align}
The three parts of the generating function ($W_r$, $W_\theta$ and $W_\phi$) increase by $2\pi J_i$ in the corresponding period $T_i$. This means that each of the $W_i$ is a periodic function, and we can do a Fourier expansion
\begin{equation}
	\frac{\partial W_i}{\partial \boldsymbol\alpha} = \sum_k A_k(\vec{J},\boldsymbol\alpha) e^{2\pi i k \theta_i} .
\end{equation}
However, the whole first term in Eq.~(\ref{eq:averageJ}) vanishes since 
\begin{equation}
\begin{split}
   \left< \dot{J}_i \right> &\approx - \frac{\dot{\alpha_m}}{T_i}\int_{T_i} \sum_{k\neq 0} 2\pi i k A_k(\vec{J},\alpha_m) e^{2\pi i k \theta_i} dt + \mathcal{O}(\dot{\alpha_m}^2,\ddot{\alpha_m}) \\ &= \mathcal{O}(\dot{\boldsymbol\alpha}^2,\ddot{\boldsymbol\alpha}),
\end{split}
\end{equation}
where we use the Einstein summation convention for the parameters $\alpha_m$. The actions are therefore invariant up to second order. \citet{Vandervoort1961} showed that the actual condition for adiabatic evolution for each of the $\alpha_m$ and each of the orbital periods $T_i$ is
\begin{equation}
	\frac{\dot{\alpha_m}}{\alpha_m} T_i \ll 1,
\end{equation}
which states that the timescale of change should be much longer than the orbital period.

\begin{figure}[!htbp]
\centering
\includegraphics[width=0.5\textwidth]{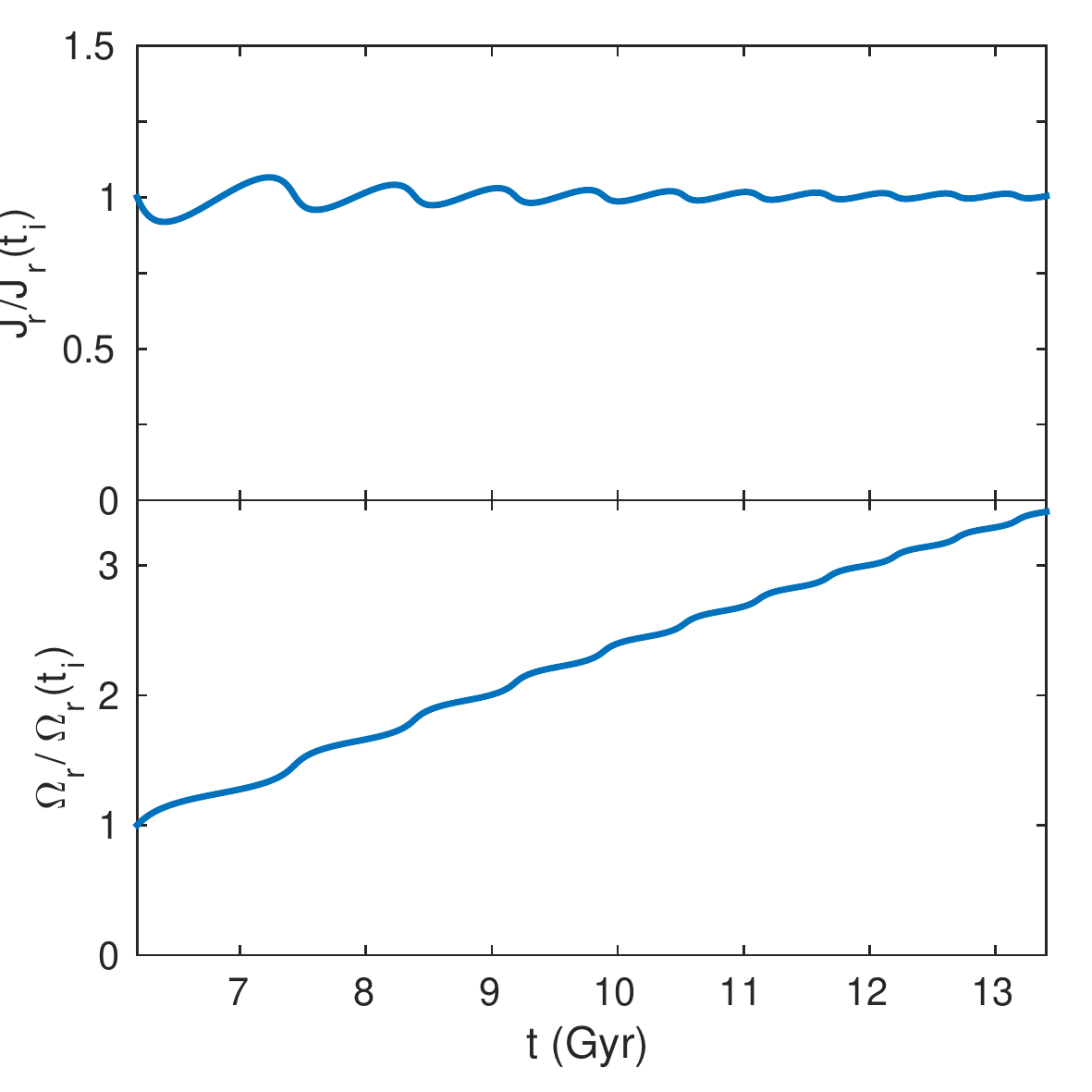}
\caption{\small Response of the radial action and the radial frequency to 
a time-dependent potential for orbit \Oon (Table \ref{tab:orbits}). 
For this orbit, the action shows a non-secular oscillation that decreases in amplitude as the 
rate of change of the potential decreases. The value of $\Omega_r$ changes significantly with time, and here there are also oscillations that become smaller as the change rate of the potential decreases. The period of the oscillations corresponds 
with the radial period. The evolution of $J_r$ has been quite adiabatic because 
it oscillates around a limiting value, which is the starting value when the 
particles were released at pericentre.}
\label{fig:ActionsFrequencies}
\end{figure}

On short timescales, $J_r$ shows a periodic oscillation that decreases in amplitude
when the changes in the parameters become smaller, as shown in
Fig.~\ref{fig:ActionsFrequencies} for one of our orbits. The final time value of $J_r$ 
returns to the initial value, indicating we are in the adiabatic regime\footnote{We 
note that if the particles were not released from pericentre, we would notice the 
initial variations in the action.}. In the case of non-adiabaticity,
 full integration of Eq.~(\ref{eq:neweqnmotion}) is needed to model the behaviour 
 of the system in action-angle space.

In an analogous way as for the actions, we compute the effect for the mean change in the angles
\begin{equation}
   \left< \dot{\theta}_i \right> \approx \left< \Omega_i(\vec{J},\boldsymbol\alpha) \right> +  \frac{\dot{\boldsymbol\alpha}}{T_i} \int_{T_i} \frac{\partial}{\partial J_i} \frac{\partial W}{\partial \boldsymbol\alpha}(\vec{q},\vec{J}, \boldsymbol\alpha) dt + \mathcal{O}(\dot{\boldsymbol\alpha}^2,\ddot{\boldsymbol\alpha}),
\end{equation}
where we can exchange the order of the derivatives
\begin{equation}
\frac{\partial}{\partial J_i} \frac{\partial W}{\partial \boldsymbol\alpha}(\vec{q},\vec{J}, \boldsymbol\alpha) =  \frac{\partial}{\partial \boldsymbol\alpha} \frac{\partial W}{\partial J_i}(\vec{q},\vec{J}, \boldsymbol\alpha) = \frac{\partial \theta_i}{\partial \boldsymbol\alpha} = 0.
\end{equation}
The result is that only the term dependent on the (evolving) frequency remains
\begin{equation}
 \left< \dot{\theta}_i \right> \approx \left<\Omega_i(\vec{J},\boldsymbol\alpha)\right> + \mathcal{O}(\dot{\boldsymbol\alpha}^2,\ddot{\boldsymbol\alpha}).
\end{equation}
The angles as a function of time may be computed from 
\begin{equation}
  \theta_i(t) \approx \theta_i(0) + \int_{0}^t \Omega_i(\vec{J},\boldsymbol\alpha(t)) dt,
\end{equation}
where like in the time-independent case, $\theta_i(0)$ is the phase at
the initial time. The frequencies $\Omega_i(t)$ are understood to be
the instantaneous frequencies at every time step. Equivalently, we can
numerically integrate the orbit in Cartesian coordinates and find
$\theta_i(t)$ from the instantaneous coordinate transformation at all
times $t$ using the generating function. The advantage of the
action-angle description is that it stresses that the phase angles
(modulo $2\pi$) depend on how the frequencies have changed with time.

\subsection{Streams in action-angle space}

\subsubsection{Sensitivity to the gravitational potential}
\begin{figure*}[!htbp]
\noindent\makebox[\textwidth]{
\includegraphics[width=0.95\textwidth]{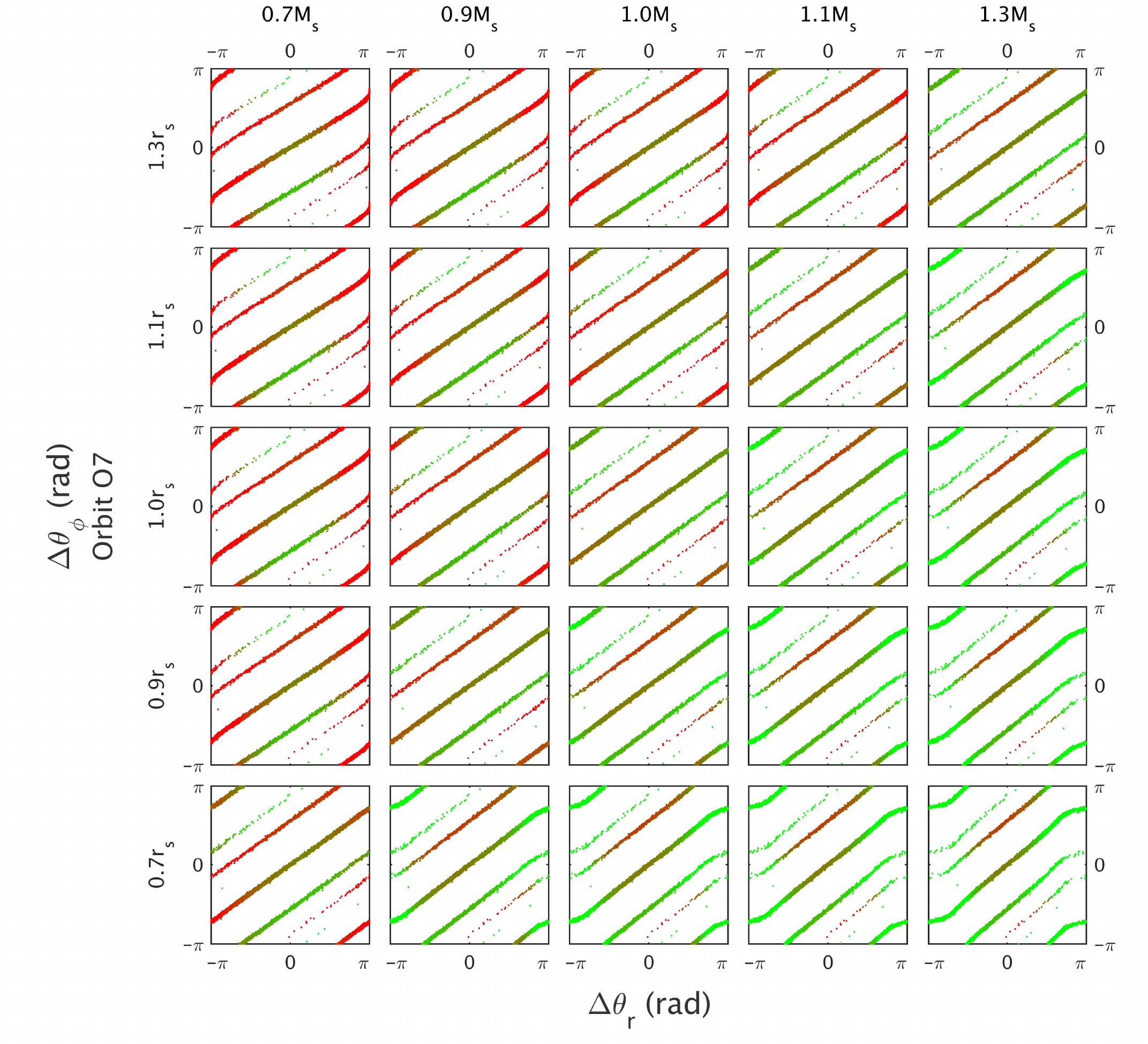}}
\vspace*{-1cm}
\caption{\small Angle space of stream \Oon (see Table \ref{tab:orbits}) evolved in a time-dependent potential, but where the angles have been computed with different (incorrect) parameters. The central panel shows the correct parameters, while $r_\textrm{s}$ is changed vertically and $M_\textrm{s}$ horizontally by 10\% and 30\%. The colour coding indicates the energies, with green representing the leading arm (more bound) and red the trailing arm (less bound).}
\label{fig:DeltaTheta}
\end{figure*}

\begin{figure*}[!htbp]
\noindent\makebox[\textwidth]{
\includegraphics[width=0.95\textwidth]{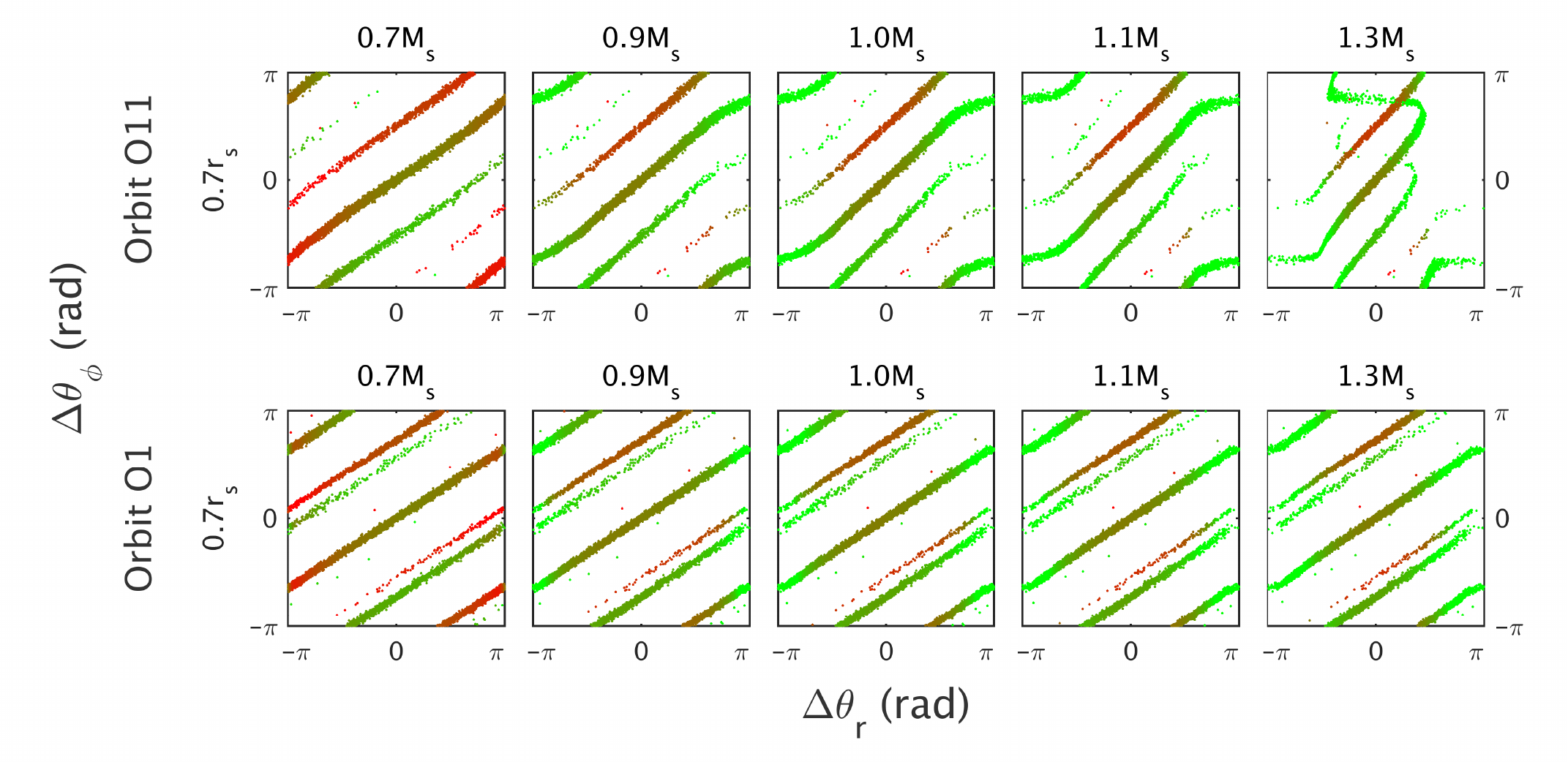}}
\vspace*{-1cm}
\caption{\small Angle space of streams \Otw (top) and stream \Oel (bottom)  evolved in a time-dependent potential, but where the angles have been computed with different (incorrect) parameters. We note that although the distortions in angle space are less pronounced for the more radial orbit \Oel, the energy gradient (indicated by the colour gradient and coding used before) is not preserved.}
\label{fig:DeltaThetaExtra}
\end{figure*}

We have mentioned that if the initial spread
in angles is negligible, then the slopes in the angle and
frequency spaces should be equal for a static potential  \citep{Sanders2013a,Sanders2013b}, as can be seen from Eq.~(\ref{eq:anglespreads}), if computed in the true potential. An example of the distortion in angle space for a stream now evolved in a time-dependent
potential where the angles (and the energy) were computed using
incorrect parameters is shown in Fig.~\ref{fig:DeltaTheta} for experiment \Oon. The
central panel corresponds to the parameters in the true (final)
potential, while we varied $r_\textrm{s}$ in the vertical and $M_\textrm{s}$ in the horizontal direction. The parameters in the eight panels around the central panel were changed by only 10\%, and even in this case, some
deviations from the expected straight lines are visible. When the
parameters were changed by as much as 30\%, as in the outer panels, the deviations become much more pronounced and are very
strong. We note that similar results are found when the stream is
evolved in a static potential.

The more incorrect the potential, the more we see wiggles in angle
space. Similar distortions are also present in the frequency spaces and
action spaces. Furthermore, the energy gradient along the stream is no
longer preserved as the parameters of the potential are varied, which
implies that the most bound (least bound) particles in the specific
trial potential are no longer `found' at the end of the leading tail
(trailing tail). Although the streams have a normal
appearance in physical space, the coordinate transformation to action-angle space and the computation of the energy are incorrect when the wrong gravitational potential is assumed. This leads to the
distorted appearance and broken energy gradient seen in Fig.~\ref{fig:DeltaTheta}. Clearly, these imprints will only be obvious for sufficiently
long streams, that is, those more extended than $2\pi$ in angle space.

We note that the distortions in angle space are less pronounced for the panels
located along the diagonal that runs from bottom left to top right in Fig.~\ref{fig:DeltaTheta}, for which 
the enclosed mass within the orbit is the same for the given $M_\textrm{s}$ and $r_\textrm{s}$. 

The degree to which these distortions and broken energy gradients manifest also depends on the type of orbit the stream progenitor has
followed. This is shown in Fig.~\ref{fig:DeltaThetaExtra} for a scale
radius that is 70\% of the true value and for different scale masses
$M_\textrm{s}$. The top panels correspond to experiment \Otw, which is on a
relatively circular orbit of apocentre-to-pericentre ratio of 2, while the
experiment in the bottom \Oel has a relatively radial orbit, with
apocentre-to-pericentre ratio of 18.

\subsubsection{Sensitivity to time-dependence}
In the time-dependent potential the exact correspondence between the
slopes of the spreads in the frequencies and in the angles is broken,
meaning that Eq.~(\ref{eq:sameSlope}) is no longer valid. This is because the
evolution of $\Delta \theta_\phi$ and $\Delta \theta_r$ depends on
integrals over time, that is,
\begin{equation}
\label{eq:angleSlopes}
\Delta
\theta_\phi / \Delta \theta_r \approx \int \Delta \Omega_\phi dt \Bigm/
\int \Delta \Omega_r dt
\end{equation} 
(assuming the initial angle spreads are negligible). 

\begin{figure}[!htbp]
\centering
\includegraphics[width=0.5\textwidth]{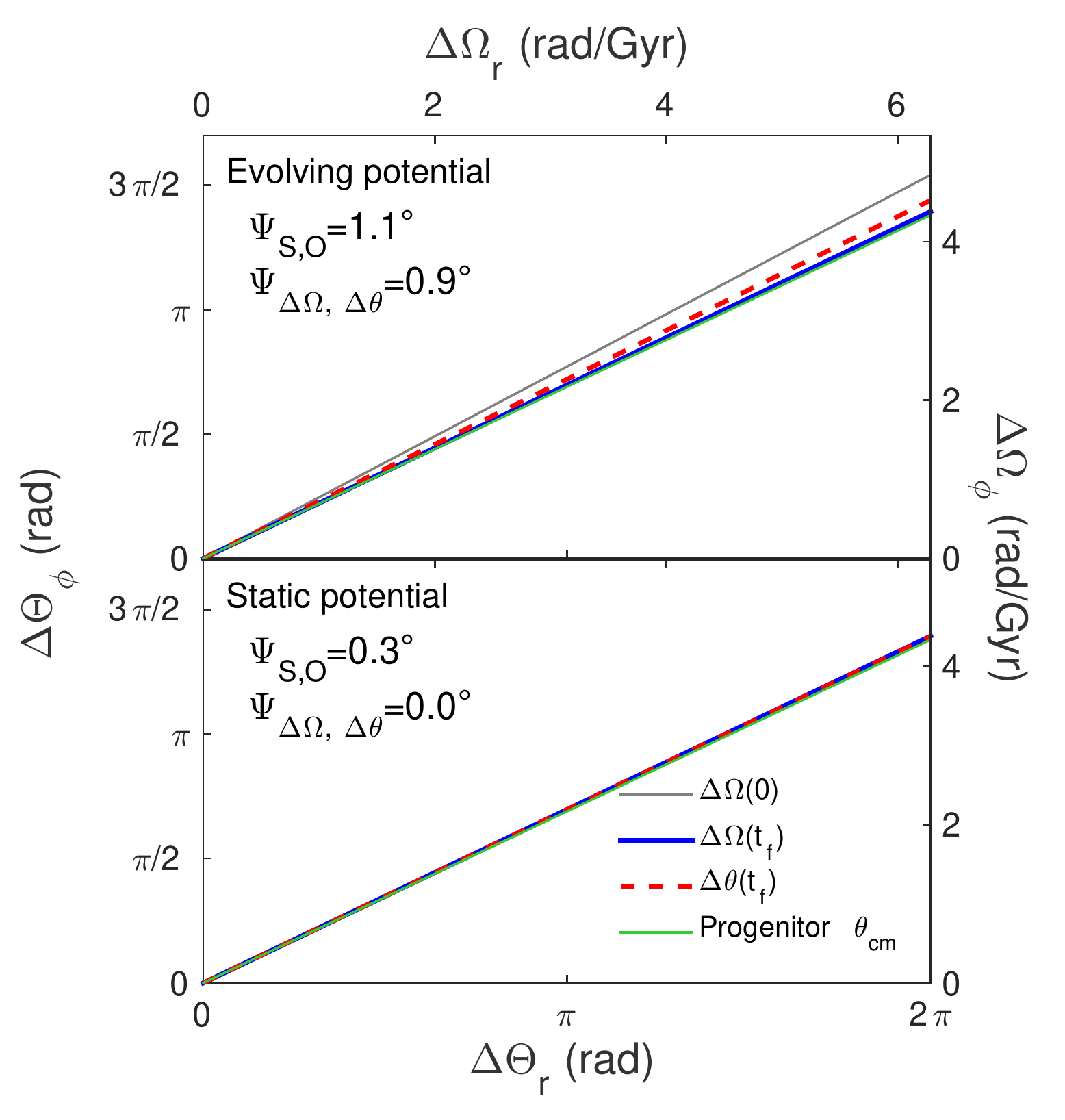}
\caption{\small Lines fitted to the angle space (red dashed line) and the frequency space (blue line) of stream \Oon evolved in a time-dependent (top) and static potential (bottom) using the same (correct) final potential. Additionally,
the green line shows the track of the progenitor assuming the potential is static (i.e.\ $\theta_\textrm{cm}(t) = \Omega_\textrm{cm}(t_f) t$), and the grey line shows the line fitted to the frequency space at the beginning of the simulation. The inset labels show the numerical values of the aperture angle (misalignment) between the stream and the (static) progenitor orbit $\Psi_\textrm{S,\,O}$, and between the stream in angle and in frequency space $\Psi_{\Delta\theta,\,\Delta\Omega}$ (see Eqs.~(\ref{eq:tangentFrequencyAngles}) and (\ref{eq:tangentstreamorbit})).}
\label{fig:slope2}
\vspace*{1cm}
\end{figure}

In Fig.~\ref{fig:slope2} we compare the lines fitted to the frequency and 
 angle spaces for the stream from orbit \Oon in the time-dependent and static cases for the correct final potential. 
 The top panel shows that for the time-dependent case, the initial slope in frequency
space (grey line) is steeper than at the end of the simulation (blue line). 
This is because the orbit shrinks as the mass increases\footnote{We recall that the ratio
of angular to radial frequencies for any gravitational potential is
limited by the homogeneous sphere case, for which $\Omega_\phi /
\Omega_r = 1/2$ and the Kepler case, for which $\Omega_\phi / \Omega_r
= 1$. As mass increases and the orbit shrinks, we may say, effectively,
that the stream moves farther away from experiencing a Kepler
potential (and hence a steeper slope) and closer to the homogeneous
sphere \citep[a shallower slope, as also seen in][]{Gomez2010a}, although
the NFW potential is neither of these limiting cases.}. In the
time-independent case we see, as expected, that both lines coincide. 
Another difference between the static and evolving case is that the stream-orbit 
misalignment changes (compare the green and red dashed lines), which is only caused by the changes in the angle space due to time evolution. 
The behaviour in angle space (indicated by the red dashed line) may be understood 
from the fact that the angles' slope is like a time average of the frequencies' slope, and is
therefore expected to lie between the initial and the final frequency
slope as observed. Therefore, a clear signature of time evolution is a
difference in the slope derived for the angles and that derived for
the frequencies, even if they are computed using the present-day
gravitational potential.

\begin{figure*}[!htbp]
\centering
\noindent\makebox[\textwidth]{
\includegraphics[width=1.0\textwidth]{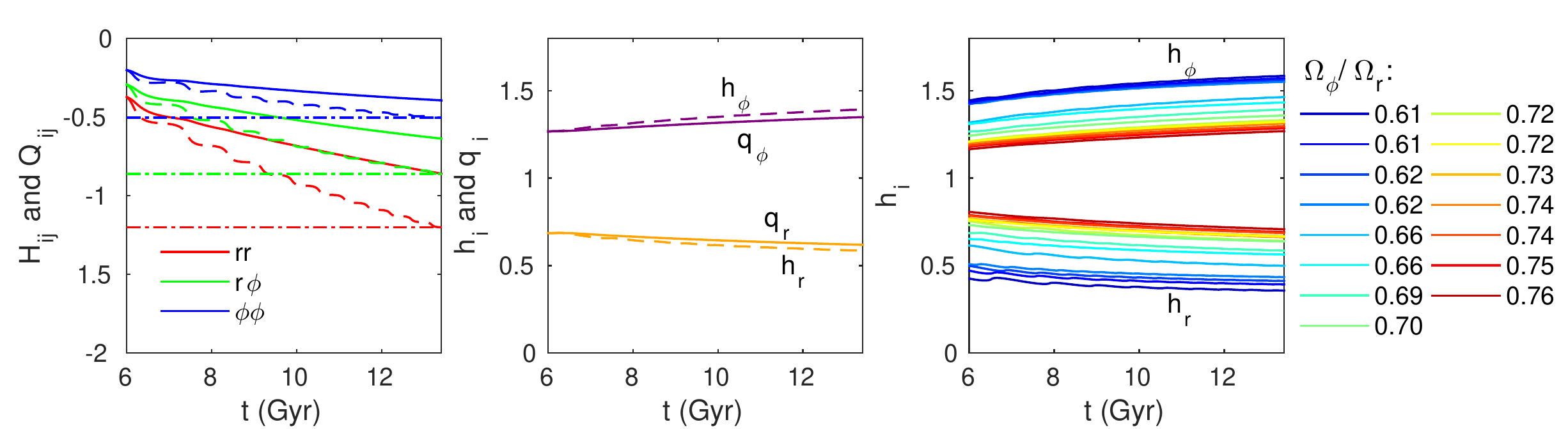}}
\caption{\small Hessians $H_{ij}$, time-averaged Hessians $Q_{ij}$ (both in units of $10^{-3}\  \textrm{kpc}^{-2}$), and their ratios $h_{i}$ and $q_{i}$ for orbit \Oon. Left panel: the dash-dotted lines show $H_{ij}$ in the static potential, the solid lines the $H_{ij}$ in the time-dependent potential, and the dashed lines $Q_{ij}$ in the time-dependent potential. We note that $|H_rr| > |H_{r\phi}| > |H_{\phi\phi}|$, and similarly for $Q_{ij}$. The middle panel shows that $h_\phi > q_\phi$ at all times and that both increase with time. On the other hand, $h_r < q_r$ at all $t$, and both decrease with time. Right panel: the ranking of the $h_i$ for the different orbits. Both $h_\phi$ and $h_r$ are ranked by the ratio $\Omega_\phi/\Omega_r$, indicating the dependence on the type of orbit. The $q_i$ are ranked in the same way.}
\label{fig:Hessians}
\end{figure*}

We now explore an analytic model to describe the
behaviour of the frequency and angle slopes more generally. We may derive the slope
in frequency space $S(\Delta\Omega)$ 
by making a Taylor expansion in the actions near the centre of mass of the progenitor \citep{Helmi1999}
\begin{equation}
  \Delta \Omega_i^k = \frac{\partial \Omega_i}{\partial J_j} \Delta J_j^k +  \mathcal{O}(\Delta {J_j^k}^2).
\label{eq:deltaOmega}
\end{equation}
The expansion is done for the $k$-th particle, all terms are evaluated at time $t$. The derivative is evaluated for the progenitor orbit and we used the Einstein summation convention on the subscript indices. We recall that 
\begin{equation}
  \frac{\partial \Omega_i}{\partial J_j} = \frac{\partial^2 H}{\partial J_i \partial J_j} \equiv H_{ij}.
\end{equation}
Working out the frequency spreads, we find
\begin{equation*}
\begin{split}
  \Delta \Omega_r^k(t)      &\approx H_{rr}(t)       \Delta J_r^k + H_{r\theta}(t)      \Delta J_\theta^k + H_{r\phi}(t) \Delta J_\phi^k,\\
  \Delta \Omega_\phi^k(t)   &\approx H_{\phi r}(t)   \Delta J_r^k + H_{\phi\theta}(t)   \Delta J_\theta^k + H_{\phi\phi}(t) \Delta J_\phi^k.
\end{split}
\end{equation*}
The cross terms are second derivatives of the Hamiltonian and
therefore symmetric, $H_{ij} = H_{ji}$. Furthermore, in a spherical
potential, the Hamiltonian is only a function of $E$ and $L$, so that
for all $j$, $H_{\theta j} = H_{\phi j}$\footnote{This is only true if $J_\phi \geq 0$, otherwise every derivative w.r.t.\ $J_\phi$ incorporates a term $\textrm{sign}(J_\phi)$ because $L = J_\theta + |J_\phi|$. Throughout this work we assume $J_\phi \geq 0$.}. The evolution of the slope
of $S(\Delta \Omega)$ can then be found from
\begin{equation}
\begin{split}
  \label{eq:DeltaOmegaEvolution}
  S(\Delta \Omega) = \frac{\Delta \Omega_\phi^k(t)}{\Delta \Omega_r^k(t)} &\approx \frac{\displaystyle \frac{\Delta J_r^k}{\Delta J_\theta^k + \Delta J_\phi^k} + \frac{H_{\phi \phi}(t)}{H_{\phi r}(t)}}{\displaystyle 1 + \frac{H_{rr}(t)}{H_{\phi r}(t)}       \frac{\Delta J_r^k}{\Delta J_\theta^k + \Delta J_\phi^k}} = \frac{R_\textrm{J}^k + h_\phi(t)}{1 + R_\textrm{J}^k h_r(t)},  \\
 {\rm where} \;\; R_\textrm{J}^k &\equiv \frac{\Delta J_r^k}{\Delta J_\theta^k + \Delta J_\phi^k} ,\\
  h_\phi(t) &\equiv \frac{H_{\phi \phi}(t)}{H_{\phi r}(t)}, \\
  h_r(t) &\equiv \frac{H_{rr}(t)}{H_{\phi r}(t)}.
\end{split}
\end{equation}
Here $R_\textrm{J}^k$ is the ratio of the actions and $h_i$ the ratio of the
Hessians. The Hessians $H_{ij}$ and the Hessian ratios $h_i$ are shown
in Fig.~\ref{fig:Hessians} and are ranked as $|H_{rr}| > |H_{\phi r}|
> |H_{\phi\phi}|$. But most importantly, the resulting ratio $h_r$
increases with time, while $h_\phi$ decreases with time. Assuming that
the action ratio $R_\textrm{J}^k$ remains constant (adiabatic limit), we find
from this equation that the slope in frequency space $S(\Delta\Omega)$ decreases with time.

For the angles of the $k$-th particle we can use the following approximation
\begin{equation}
\begin{split}
  \Delta \theta_i^k(t) &= \Delta \theta_i^k(0) + \int_0^t \Delta\Omega_i^k(t) dt \\
   &= \Delta \theta_i^k(0) + \int_{0}^{t} H_{ij}(t) \Delta J_j^k(t) dt +  \mathcal{O}(\Delta {J_i^k}^2)  \\
                            &\approx \int_{0}^{t} H_{ij}(t) \Delta J_j^k(t) dt \ \\
                            &\approx \Delta J_j^k \int_{0}^{t} H_{ij}(t) dt ,
\end{split}
\label{eq:anglespreads2}
\end{equation}
where we have neglected the initial angle spread and in the last approximation assumed adiabaticity of the
actions. Therefore $\Delta J_j^k$ may be evaluated at any time step,
although we typically use the values at the final time in our
computations.

We now focus on the slope in angle space. If we define $\tfrac{1}{t} \int_{0}^{t} H_{ij} dt \equiv Q_{ij}$ (the time-averaged Hessian), then
\begin{equation*}
\begin{split}
  \Delta \Theta_r^k(t) &\approx   \left( Q_{rr}(t)       \Delta J_r^k + Q_{r\theta}(t)      \Delta J_\theta^k + Q_{r\phi}(t) \Delta J_\phi^k\right) t ,\\
  \Delta \Theta_\phi(t) &\approx   \left( Q_{\phi r}(t)   \Delta J_r^k + Q_{\phi\theta}(t)   \Delta J_\theta^k + Q_{\phi\phi}(t) \Delta J_\phi^k \right) t .
\end{split}
\end{equation*}
The slope in the angles can then be found from
\begin{equation}
\begin{split}
  \label{eq:DeltaThetaEvolution}
  S(\Delta \theta) = \frac{\Delta \Theta_\phi(t)}{\Delta \Theta_r(t)} &\approx \frac{  \displaystyle  R_\textrm{J}^k + q_\phi(t) }{\displaystyle 1 + q_r(t) R_\textrm{J}^k}, \\
   {\rm where} \;\; q_\phi(t) &\equiv \frac{Q_{\phi \phi}(t)}{Q_{\phi r}(t)}, \\
  q_r(t) &\equiv \frac{Q_{rr}(t)}{Q_{\phi r}(t)}.
\end{split}
\end{equation}
Here the $q_i$ are the ratios of the $Q_{ij}$. The ranking of the $Q_{ij}$
and the $q_i$ behaves very similarly to that of the $H_{ij}$ and $h_i$,
as shown in Fig.~\ref{fig:Hessians}. In the same manner as for the
frequency slope, we infer that $S(\Delta\theta)$ is also a
decreasing function of time.

We are now ready to compute the difference between $S(\Delta\theta)$
and $S(\Delta\Omega)$ by using that the difference between the $q_i$
and the $h_i$ is small:
\begin{align}
	q_r(t) &\equiv h_r(t) - \epsilon_r(t), \\
	q_\phi(t) &\equiv h_\phi(t) + \epsilon_\phi(t),
\end{align}
where the appropriate sign was chosen such that the $\epsilon_i$ are
always positive. We define for computational ease
\begin{equation}
\begin{split}
	T(\Delta\Omega) &= \frac{1}{S(\Delta\Omega)}, \\
	T(\Delta\theta) &= \frac{1}{S(\Delta\theta)}.
\end{split}
\end{equation}
We may expand the inverse of the angle slope $T(\Delta\theta)$ 
\begin{equation}
\begin{split}
	T(\Delta\theta) &= \frac{1 + R_\textrm{J}^k q_r}{R_\textrm{J}^k + q_\phi} = \frac{1 + R_\textrm{J}^k h_r - R_\textrm{J}^k \epsilon_r}{R_\textrm{J}^k + h_\phi + \epsilon_\phi} \\
	          &= \frac{1 + R_\textrm{J}^k h_r - R_\textrm{J}^k \epsilon_r}{R_\textrm{J}^k + h_\phi} \left(\frac{R_\textrm{J}^k + h_\phi}{R_\textrm{J}^k + h_\phi + \epsilon_\phi} \right) \\
	          &\approx \frac{1 + R_\textrm{J}^k h_r - R_\textrm{J}^k \epsilon_r}{R_\textrm{J}^k + h_\phi} \left(1 - \frac{\epsilon_\phi}{R_\textrm{J}^k + h_\phi} \right) \\
	          &\approx \frac{1 + R_\textrm{J}^k h_r - R_\textrm{J}^k \epsilon_r}{R_\textrm{J}^k + h_\phi} - \epsilon_\phi \frac{1 + R_\textrm{J}^k h_r }{(R_\textrm{J}^k + h_\phi)^2}.
\end{split}
\end{equation}
Using Eq.~(\ref{eq:DeltaOmegaEvolution}) to compute $T(\Delta\theta)$,
we can derive the difference of the inverse slopes
\begin{equation}
\begin{split}
	T(\Delta\theta) - T(\Delta\Omega) &= \frac{S(\Delta\Omega)-S(\Delta\theta)}{S(\Delta\theta) S(\Delta\Omega)}  \\
	&= -\frac{R_\textrm{J}^k \epsilon_r}{R_\textrm{J}^k + h_\phi} - \epsilon_\phi \frac{1 + R_\textrm{J}^k h_r }{R_\textrm{J}^k + h_\phi}  \\
	&= -\frac{T(\Delta\Omega) \epsilon_\phi + R_\textrm{J}^k \epsilon_r}{R_\textrm{J}^k + h_\phi} < 0.
\end{split}
\end{equation}
All the terms in the fraction are positive, and $T(\Delta\Omega)$ is
therefore always greater than $T(\Delta\theta)$. This means that
$S(\Delta\theta)$ is always larger than $S(\Delta\Omega)$, and this
difference will increase with time. The longer the system is evolved
in a time-dependent potential, the larger the difference in slopes
becomes with respect to the same final static
potential. This also means that streams that fell in at early times
will show a stronger signature of the separation between the angle and
frequency relations than streams that fell in more recently.

\begin{figure}[!htbp]
\centering
\includegraphics[width=0.5\textwidth]{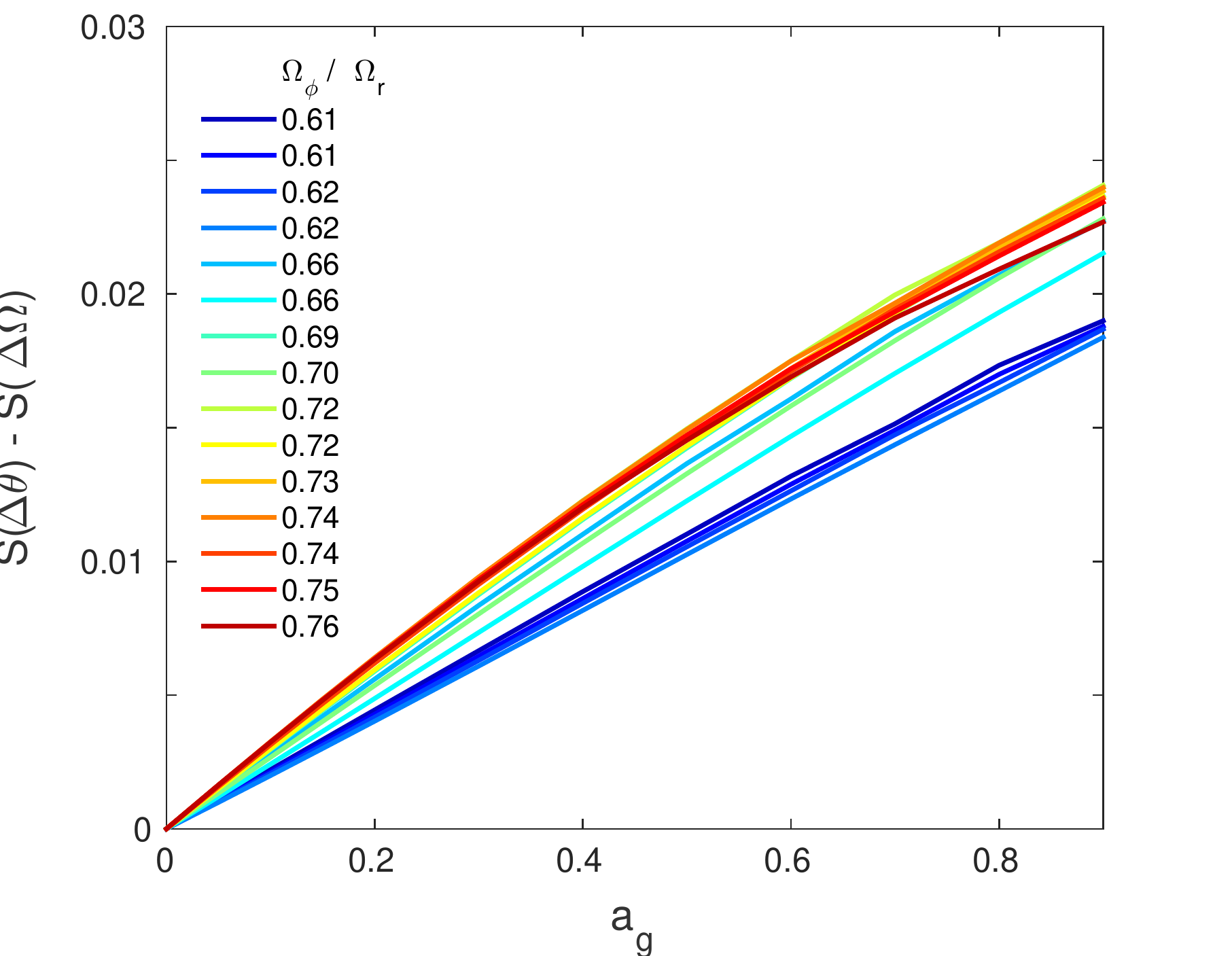}
\caption{\small Slope differences between the streams in angle and in
  frequency space as a function of growth parameter $a_\textrm{g}$, as
  predicted by our analytic model using Eq.~(\ref{eq:deltaOmega}) and (\ref{eq:anglespreads2})
  for our experiments.}
\label{fig:AgSlope}
\end{figure}

In Fig.~\ref{fig:AgSlope} we show the predictions of the slope
differences for the angle and frequency spaces for the orbits in Table~\ref{tab:orbits}. With increasing amount of time evolution by changing the growth parameter $a_\textrm{g}$, the
difference in slope also becomes larger. Another interesting prediction is
that the more circular orbits have a larger slope difference, although this is at a small level. We gauge the magnitude of the angle-frequency misalignment by using the aperture angle $\Psi_{\Delta\theta,\,\Delta\Omega}$ between the fitted lines in the angle and frequency space (as also used by \citet{Sanders2013a, Sanders2013b})
\begin{equation}
	\Psi_{\Delta\theta,\,\Delta\Omega} = \Psi_{\Delta\theta} - \Psi_{\Delta\Omega} \approx \tan\left( \Psi_{\Delta\theta} - \Psi_{\Delta\Omega} \right) = \frac{S(\Delta\theta) - S(\Delta\Omega)}{1 + S(\Delta\theta)S(\Delta\Omega)},
\label{eq:tangentFrequencyAngles}
\end{equation}
where $S(i) = \tan(\Psi_i)$, and we have used that the difference in slopes is small\footnote{In these equations we have ignored the slope in the $r$-$\vartheta$ angle and frequency spaces, because the slopes are very close to those of the $r$-$\phi$ spaces.}. With this definition of the angle-frequency misalignment, typical values of $\Psi_{\Delta\theta,\,\Delta\Omega}$ are around $1.5^\circ$ for $a_\textrm{g}=0.8$. 

It has been discussed earlier that a stream does not exactly follow an orbit, and the small difference between these two trajectories is known as the stream-orbit misalignment. The characteristic magnitude of this can be found from
\begin{equation}
	\Psi_\textrm{S,\,O} = \Psi_{\Delta\theta} - \Psi_{\theta}\approx \tan\left( \Psi_{\Delta\theta} - \Psi_{\theta_\textrm{cm}}\right) = \frac{S(\Delta\theta) - S(\theta_\textrm{cm})}{1 + S(\Delta\theta)S(\theta_\textrm{cm})},
\label{eq:tangentstreamorbit}
\end{equation}
where $S(\theta_\textrm{cm}) = \tan{\Psi_{\theta_\textrm{cm}}}$ is the slope of the
straight line traced by the centre of the mass of the progenitor in
angle space for a static potential (see also
Fig.~\ref{fig:slope2}). The typical order of magnitude of $\Psi_\textrm{S,\,O}$
for the Carina progenitor is about 1 degree, similar to what was found
in \citet{Sanders2013a} for a static logarithmic axisymmetric
potential, and also similar to values of the stream-orbit
misalignment in the isochrone potential of
\citet{Eyre2011}. We note that the angle-frequency differences in slope that we find are of the
same order of magnitude as the stream-orbit misalignment.
\begin{figure*}[!htbp]
\noindent\makebox[\textwidth]{
\includegraphics[width=0.7\textwidth]{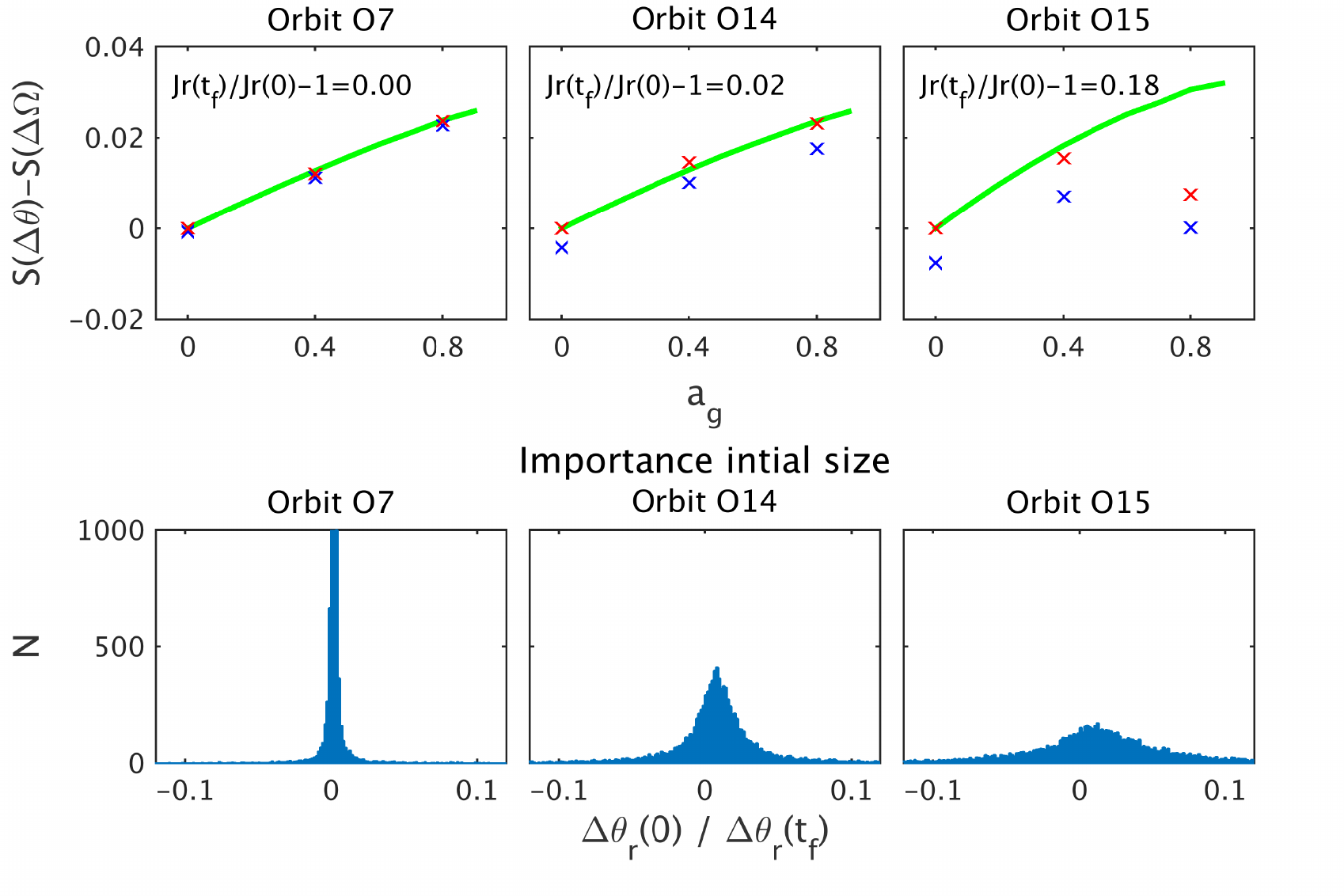}}
\vspace*{-0.7cm}
\caption{\small Examples of the slope differences between the angle and frequency distributions (top panels) and histograms of the spread of $\Delta \theta_r(0)/\Delta \theta_r(t_f)$ (bottom panels). In the top panels, the blue crosses indicate values from the full simulation, while the green lines are the predictions from our analytic model. The red crosses are the slopes measured when the initial spreads from the simulations are removed. All orbits now align with the model when $a_\textrm{g}=0$. For large $a_\textrm{g}$ the more circular orbits such as orbit \Oei deviate strongly from the model because of non-adiabaticity.}
\label{fig:acModels}
\end{figure*}

When comparing simulations with and without time evolution, the
stream-orbit misalignment can also be used as an indicator of
time evolution. In Fig.~\ref{fig:xyexample} we saw that the angular
position of the stream and progenitor orbit petals one radial period away from the current progenitor location are different for
the time-dependent case. This angular separation can be derived as
follows. For the progenitor, the location of the petal is simply the
precession of the orbit in one radial period, that is, $\psi_\textrm{cm} = 2\pi
S(\theta_\textrm{cm}) = 2 \pi \Omega_\phi/\Omega_r$. For the stream, this
difference in radial angle from the progenitor is $\Delta \Theta_r =
2\pi$, and therefore the angle of the petal of the stream is
$\psi_\text{stream} = 2\pi S(\Delta\theta)$. Therefore the stream-orbit misalignment in
azimuthal angle is
\begin{equation}
	\Delta \phi_\textrm{S,\,O} \approx \psi_\text{stream} - \psi_\textrm{cm} = 2\pi\left( S\left(\Delta\theta\right) - S\left(\theta_\textrm{cm}\right)\right) .
\label{eq:streamorbitmisalignmentazimuthal}
\end{equation}
Increasingly higher growth rates of the potential will yield also
larger angular stream-orbit separations because $S(\Delta\theta)$ changes 
by the amount in Fig.~\ref{fig:AgSlope} ($S(\Delta\Omega)$ and $S(\theta_\textrm{cm})$ are 
by construction the same at the final time). By multiplying this by $360^\circ$ to 
find the difference at the first petals next to the progenitor, we find that 
these differences can reach up to $10$ degrees in our experiments.

In summary, stream particles are distributed following straight lines
in angle space when computed using the correct potential and
otherwise show a wiggly behaviour and a broken energy gradient. These straight lines have the
same slope in frequency space in the static case. If the
potential has evolved in time, then the slope in frequency space is
typically shallower than in angle space, to an extent that depends on
the growth of the gravitational potential.

\subsubsection{Validation of the analytic model}
We expect our analytic model to faithfully reproduce the behaviour of the
test-particle simulations except in some cases. We neglected two effects that
can modify these slopes in the derivations
above. The first appears when the initial angle spread of the
progenitor is comparable to its current extent, in which case the
former needs to be taken into account. Since the initial extent typically is unknown, it is better to use very extended streams to derive the evolution in time of the host potential. The second
effect is when the time evolution of the potential is
non-adiabatic. In this case, the higher order terms of
Eq.~(\ref{eq:neweqnmotion}) have to be taken into account.

To confirm this, we used three simulations with different growth
factors $a_\textrm{g} = \{ 0.0, 0.4, 0.8\}$, but with the same initial action
distribution as in the $a_\textrm{g}=0.8$ case by using the
transformation described in Eq.~(\ref{eq:lintransformation}). 

\begin{figure*}[!htbp]
\centering
\noindent\makebox[\textwidth]{
\includegraphics[width=1.\textwidth]{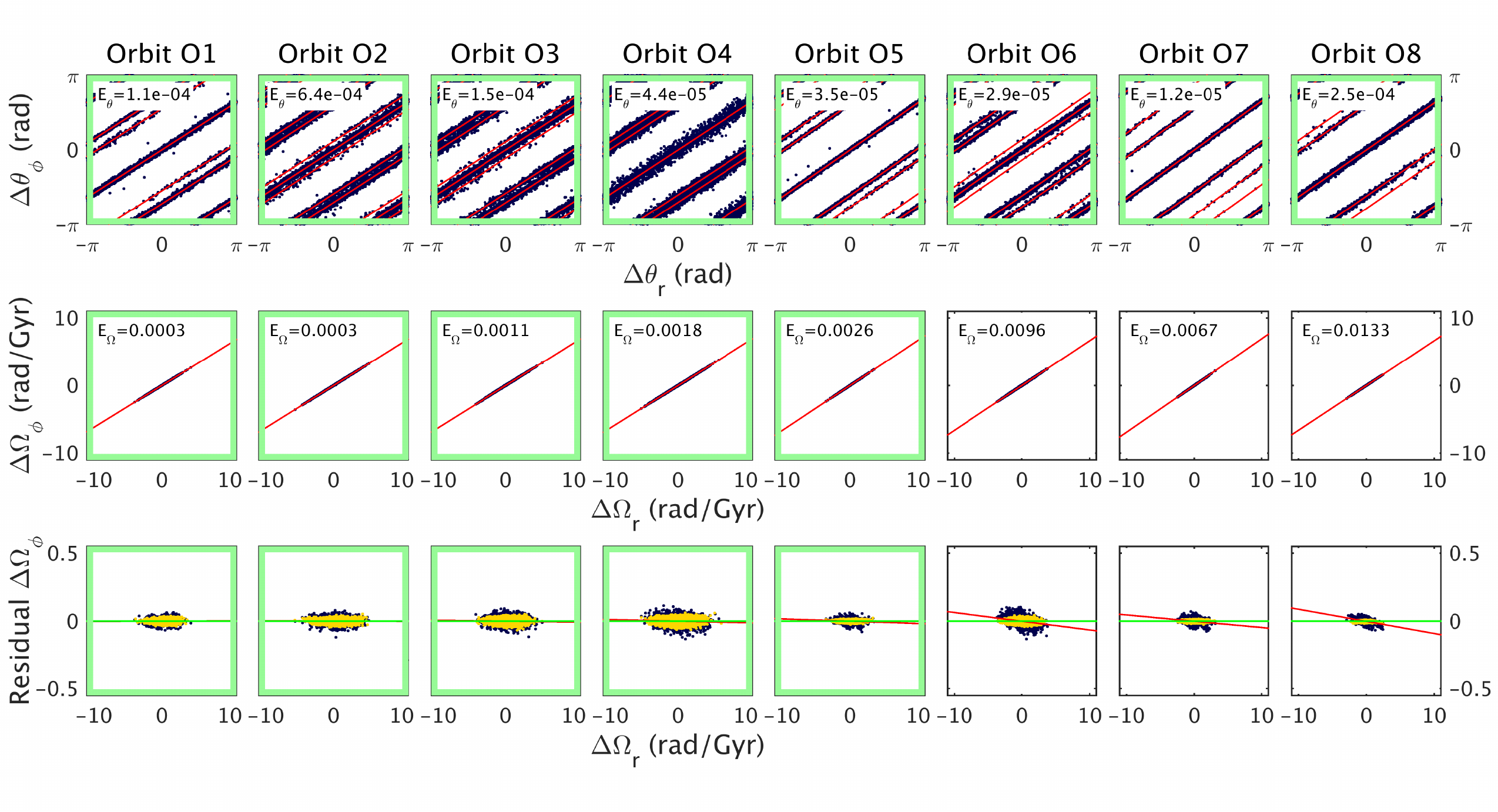}}
\vspace*{0.0cm}
\noindent\makebox[\textwidth]{
\includegraphics[width=1.\textwidth]{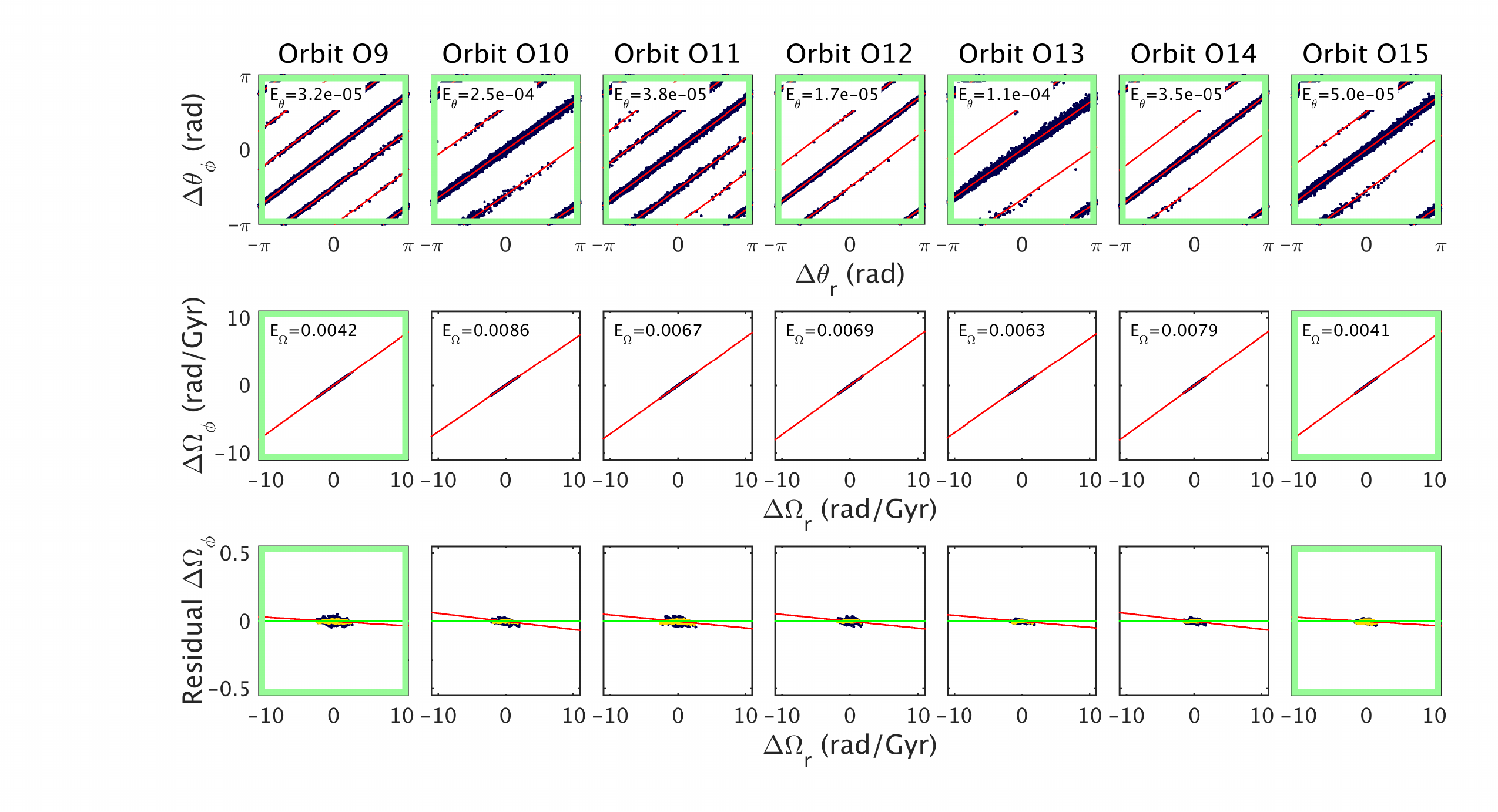}}
\vspace*{-1cm}
\caption{\small Angle and frequency space plots and residuals of the
  frequency space fits of all 15 streams for $1\%$ distance (and proper motion) errors in
  the time-dependent potential. The streams are centred on the
  position of the centre of mass of the progenitor in the error-free
  case. The black points are error-convolved, while those in yellow are
  error-free and only shown in the bottom row. The residuals were
  computed by subtracting the best-fit straight line to the error-free
  frequency distribution (green line) from $\Delta\Omega_\phi$. The
  text in the panels shows the relative difference in the slopes for
  the error-convolved and error-free cases derived in angle space
  $E_\theta$ and in frequency space $E_\Omega$ (i.e.\ using the red vs green straight lines, respectively). The panels for which
  $E_i < 0.005$ are marked with a green box. The difference in
  slope in angle space clearly is generally far below this value, while the
  frequencies are much more strongly affected by the errors.}
\label{fig:withErrors}
\end{figure*}

Figure~\ref{fig:acModels} shows the difference between the slopes in
angle and in frequency spaces for our analytic model. In this figure,
the green line shows the prediction of our model using
Eqs.~(\ref{eq:DeltaOmegaEvolution}) and
(\ref{eq:DeltaThetaEvolution}),  and the actual measured
slope differences from the test-particle simulations are shown as blue crosses. This comparison shows that the agreement is excellent, as exemplified for
Orbit \Oon in the left panel, and this is true for most of the orbits
we explored. However, there are a few cases, such as those shown in the
panels on the right, where the model fails.

For these two cases, the initial extent of the progenitor in
angle space cannot be neglected, as can be seen from the bottom
panels. This can either be because the streams are relatively short
or because the initial spread in angles is large.  When the initial
angle spreads as given by Eqs.~(\ref{eq:anglespreads}) and
 (\ref{eq:anglespreads2})  are removed, the simulations match the
model significantly better, as indicated by the red crosses.

The rightmost panel still shows some disagreement, especially for
higher values of the growth factor $a_\textrm{g}$. This behaviour stems from
the non-adiabatic evolution of the orbit, so that higher terms in
Eq.~(\ref{eq:higherorderterms}), and the last approximation in
Eq.~(\ref{eq:anglespreads2}), fail. An indication that these
orbits evolved non-adiabatically is for example that $J_r(0)/J_r(t_f)-1 \approx 0.18$ for orbit \Oei. The orbits that are more circular typically suffer more strongly from this effect.  This implies that radial orbits may be preferred to obtain the amount of time evolution for a stream that evolved in a smoothly growing potential such as ours.

\section{Observational prospects}
\label{sec:ObservationalProspects}
\begin{figure*}[!htbp]
\centering
\noindent\makebox[\textwidth]{
\includegraphics[width=1.\textwidth]{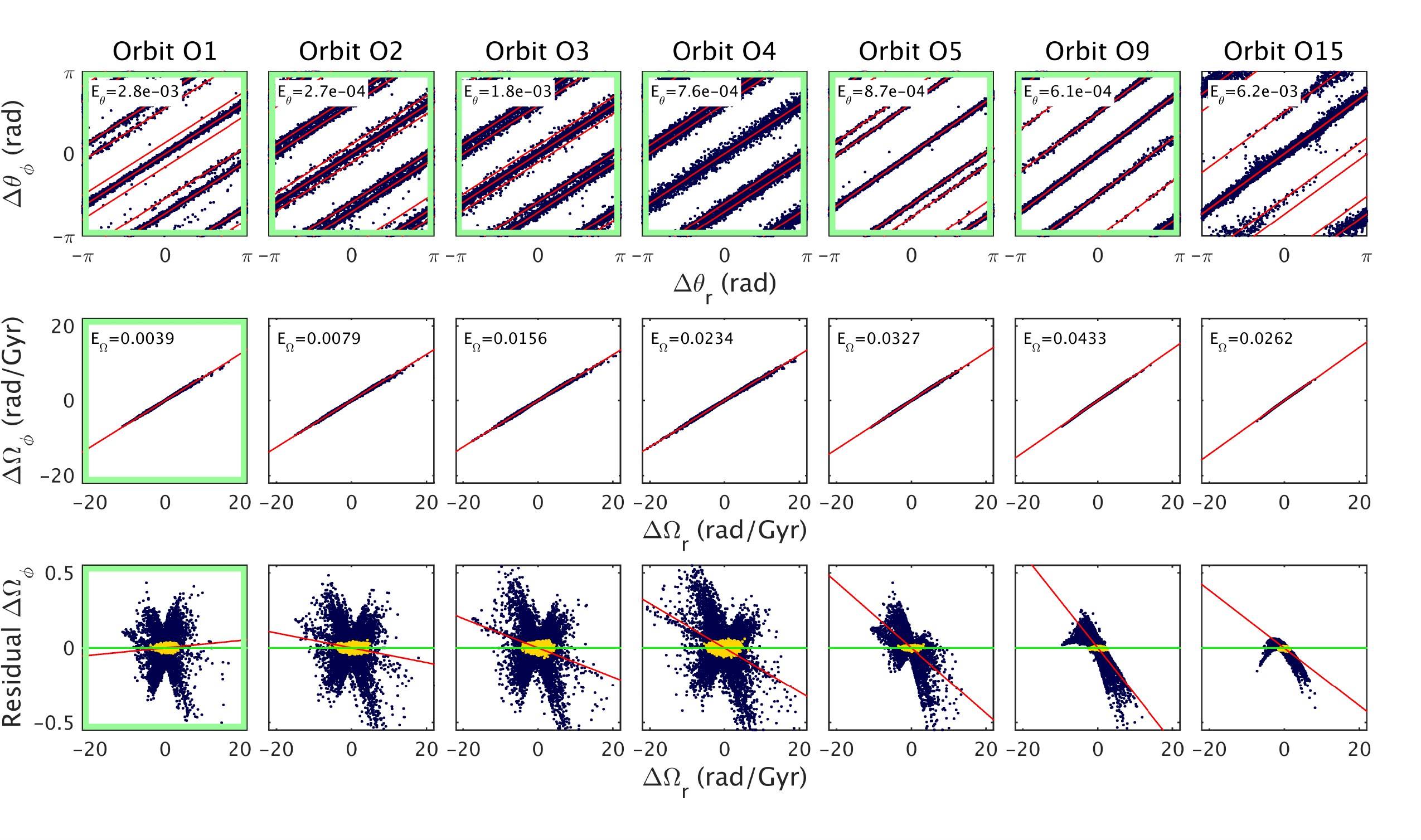}}
\vspace*{-1cm}
\caption{\small Selection of the seven best orbits of
  Fig.~\ref{fig:withErrors}, but now with $10\%$ parallax (and proper
  motion) errors. The colours, lines, and insets are the same as in
  Fig.~\ref{fig:withErrors}. With these 10$\times$ larger distance errors,
  the angle slopes of the streams are still well measured for most
  orbits, while the frequency slopes generally are not.}
\label{fig:withErrorsLarge}
\end{figure*}
\begin{figure*}[!htbp]
\noindent\makebox[\textwidth]{
\includegraphics[width=1.\textwidth]{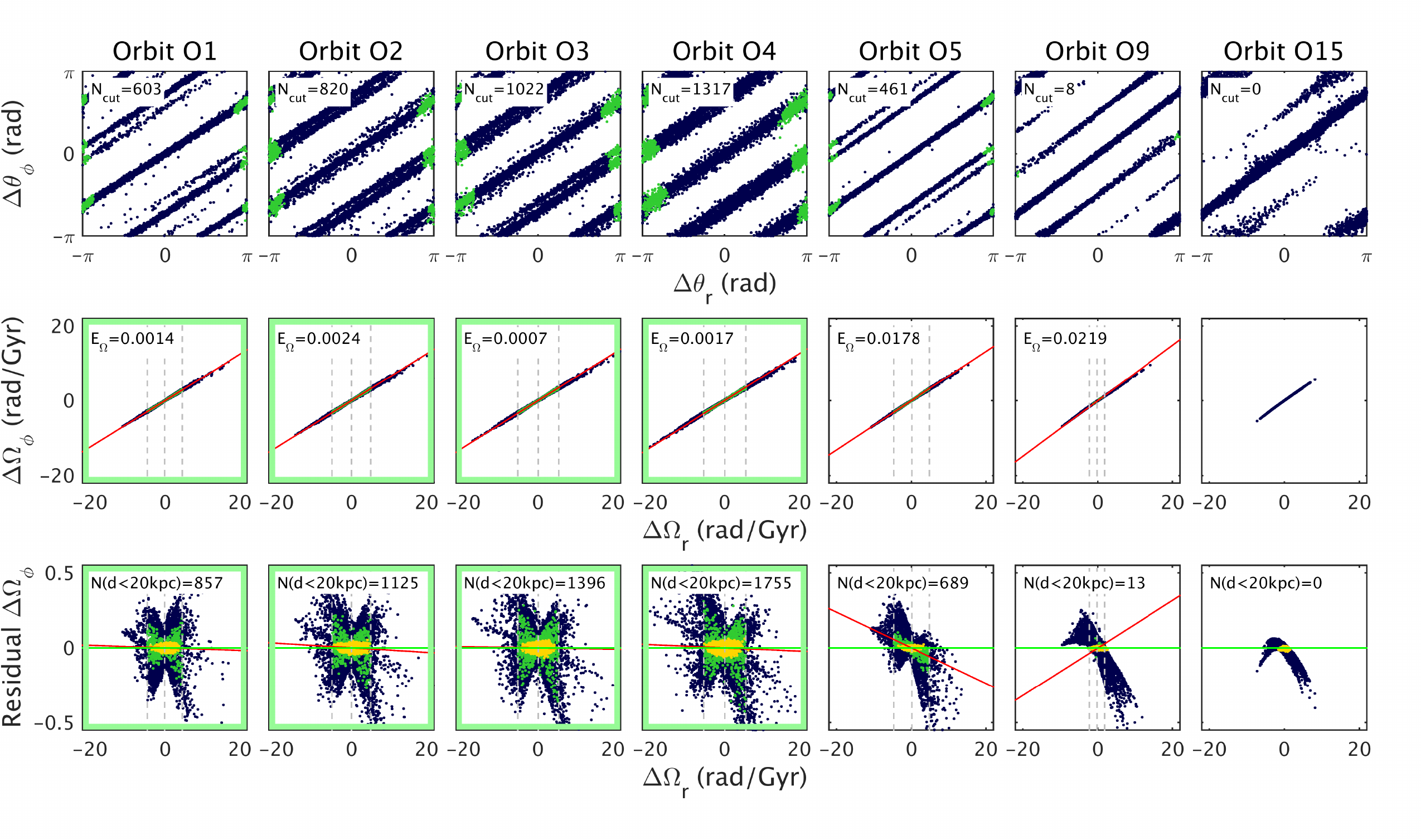}}
\vspace*{-1cm}
\caption{\small Selection of the orbits of Fig.~\ref{fig:withErrors}
  with $10\%$ parallax errors, but now with two cuts: $\sigma_d \leq
  2$ kpc and $|\Omega_r - \left< \Omega_r \right>| \leq \textrm{
    SD}_{\Omega_r}$. The colours are the same as in
  Fig.~\ref{fig:withErrorsLarge}, with the addition that green points represent the particles that remain after the cuts. The
  grey vertical dashed lines indicate $\left< \Omega_r \right>$ and $\left< \Omega_r \right> \pm \textrm{
    SD}_{\Omega_r}$  applied after the distance cut. The insets in the top panels show the number of
  remaining particles after the cuts $N_{cut}$, while those in the bottom panels are the number
  of particles within a distance of 20 kpc from the Sun $N(d<20~\textrm{kpc})$.}
\label{fig:withErrorsCuts}
\end{figure*}

Now that we have determined the signature of time evolution of a
gravitational potential through the difference in slopes in the angle
and frequency spaces, we are interested in establishing whether this
effect is measurable. This is particularly relevant in the context of
the upcoming catalogues from the Gaia satellite.

We therefore convolved our simulated data in observable space with
errors. The largest uncertainties typically come from the errors in
distance and proper motion. We investigated two cases for the errors in
the parallax: $1\%$ ($\sigma_\pi/\pi = 0.01$) and $10\%$
($\sigma_\pi/\pi = 0.1$), while the proper motion errors are set to be
$\sigma_\mu = 0.5 \sigma_\pi$), as given by the Gaia mission error
estimates. We assumed radial velocity errors of $1$ km/s, which is the
level obtainable from follow-up surveys of the Gaia satellite such as
4MOST \citep{DeJong2012} and WEAVE \citep{Dalton2012}. Although this
error may be seen as small, we have found no significant dependence on
the radial velocity error, even if this was as large as 10 km/s.
After applying error convolution, we converted the data back to
Cartesian coordinates.

In Fig.~\ref{fig:withErrors} we show the structure in angle (top row)
and in frequency (middle row) space for the $1\%$ distance errors for
all 15 orbits evolved in the time-dependent potential. The bottom row
panels show the residuals in frequency space after subtracting the
fitted slope from the error-free case (green line).  Here the
black points correspond to the error-convolved case, while the yellow points
are error free. 

We define the relative difference
in slope in the frequency $E_\Omega$ and angle $E_\theta$ spaces after
error convolution as
\begin{equation}
\begin{split}
	E_\Omega &= \left|\frac{\tilde{S}(\Delta\Omega)-S(\Delta\Omega)}{S(\Delta\Omega)}\right| , \\
	E_\theta &= \left|\frac{\tilde{S}(\Delta\theta)-S(\Delta\theta)}{S(\Delta\theta)}\right| ,
\end{split}
\end{equation}
with the $\tilde{S}$ representing the slopes derived by fitting the
distributions with errors. The various panels
in Fig.~\ref{fig:withErrors} show that this relative difference
typically varies from $10^{-5}$ up to $6 \times 10^{-4}$ for $E_\theta$ and between $3 \times 10^{-4}$ up to $10^{-2}$ for $E_\Omega$, indicating
that errors affect the distribution of frequencies more severely than
that of angles.

An estimate of tolerable errors may be derived from the following arguments. If $\Delta S$ is the slope difference between frequency and angle space (typically a few \%), then we can tolerate 
\begin{equation}
	\sqrt{E_\Omega^2 + E_\theta^2} \le \Delta S/S \approx 0.005 .
\end{equation}
The panels that satisfy this requirement are highlighted with a green
box. The very radial orbits that have small
pericentres are mostly selected in this way. These streams have many stars
relatively near the Sun (which is positioned at $8$ kpc from the
Galactic centre along the $x$ axis). The more radial streams also have a much broader spread in $J_r$, which is slightly less affected by the
errors than the more circular orbits, which have a broader spread in
angular momentum. We conclude that the $1\%$ distance errors allow us
to uncover the time evolution for at least half of our orbits when the
growth factor is $a_\textrm{g}=0.8$. However, if we were to consider a growth
factor $a_\textrm{g}=0.4$, the misalignment between angle and
frequency space would typically be reduced by a factor 2, and the errors would considerably hinder measuring the time evolution for more slowly evolving
potentials.

In Fig.~\ref{fig:withErrorsLarge} we show the seven best orbits from Fig.~\ref{fig:withErrors}, but now with $10\%$ distance
errors. The colours and insets are the same as in
Fig.~\ref{fig:withErrors}. As the insets and green boxes show, the
angles are still quite well measured, but the frequency slope is not. Only one very radial orbit (that comes close to the
Sun) presents an acceptable level of uncertainty. There is significant
structure in the bottom panels, which depict the residuals between the
distribution in frequency space before and after error
convolution. For each experiment this is a result of superposing
individual wraps with varying distance gradients, and this, after a distance-dependent error convolution,
causes different frequency distributions. The bow-tie-like
structure is thus the result of the overlap of multiple wraps
that are differently distorted in frequency space. The more circular orbits \Oei
and \Ote only show one leg of these bow-tie-like structures because
basically just slightly more than one wrap is present for these experiments. 

Given that the 10\% errors make it challenging to fit a slope in
frequency space, but less so in angle space, we
investigated methods for cutting the data and retaining the highest quality
measurements. Figure~\ref{fig:withErrorsCuts} shows the result of
combining a distance cut and a frequency cut for the same seven orbits as in
Fig.~\ref{fig:withErrorsLarge}. The cuts are given by
\begin{equation}
\begin{split}
\sigma_d &\leq 2 \text{ kpc} , \\
|\Omega_r - \left< \Omega_r \right>| &\leq \textrm{ SD}_{\Omega_r} ,
\end{split}
\end{equation}
with $\left< \Omega_r \right>$ the mean frequency from the data and
$\textrm{ SD}_{\Omega_r}$ the standard deviation of $\Omega_r$, both
computed after the distance cut. The distance error cut improves both
the tangential velocities and the distances, and its effect is to
mainly select particles at pericentre, as seen in the top panels of
this figure. Our cut at $2$ kpc essentially removes particles with
errors larger than 10\% at 20 kpc. The motivation for imposing a
frequency cut is to remove outliers. The vertical dashed lines in the
frequency space of Fig.~\ref{fig:withErrorsCuts} show the mean and
standard deviation of $\Delta\Omega_r$ with errors, and they encompass the error-free $\Delta\Omega_r$ distribution indicated in yellow quite well. The combination of these cuts still leaves many particles in the stream, as indicated by the insets in the
figure, except for orbits \Oei and \Ote, which basically have no or too
few particles closer than 20 kpc, and hence have distance errors
larger than 2 kpc. In conclusion, a cut in frequency and distance
leads to satisfactory improvements, except when there are too few
particles and for a few pathological cases. Such a cut might therefore
be used to measure the slope in frequency space, but is not necessary
in angle space.

\section{Discussion and conclusions}
\label{sec:conclusions}

We have investigated the effect of an adiabatically growing
time-dependent potential on streams. To this end we used the
inside-out growing spherical NFW potential from \citet{BuistHelmi2014}
as a background for modelling the evolution of a set of 15 test-particle
streams. 

We performed a series of numerical experiments starting from
different initial conditions that were mostly ran in a quite strongly growing
model in which the enclosed mass within the orbits that we explored
approximately doubled. These experiments showed that the precession rate
of streams, that is, the angular location of the `apocentres' of streams
in the orbital plane, is significantly different when the potential
has evolved in time. Time evolution typically leads to a misalignment
or angular difference of $\sim 10^\circ$ in comparison to the static
case, and this is roughly independent of the progenitor size. To be
able to detect this effect, however, streams need to be sufficiently
long, meaning that they need to have wrapped around more than once. 

We then analysed the behaviour of streams in action-angle
coordinates. Streams typically appear as extended linear structures in
angle and frequency space, also in the time-dependent case. We found
that time evolution causes these distributions to differ in slope,
unlike what is expected for a static potential. To explain these
findings, we developed an analytic description of a stream in an
adiabatically changing time-dependent potential.  This allowed us to
predict that the difference in slopes in angle and in frequency space
is of order $\sim 0.005 - 0.025$, that is, $0.3^\circ - 1.5^\circ$, for potentials with a growth factor ranging from $0.2$ to
$0.8$. Experiments for which the time evolution signal is most
reliably recovered were those that initially had a small size and
evolved in an adiabatic way. This condition is most easily
satisfied for the more radial orbits.

Although the predicted effect is small, we explored whether it
would be observable with the next-generation facilities for mapping the
Galaxy, such as the Gaia satellite mission \citep{prusti2012} and the follow-up surveys
4MOST \citep{DeJong2012} and WEAVE \citep{Dalton2012}. When assuming an error in
parallax of 1\% (and half of that in the proper motion), we found that
for about half of the streams the time-dependence signature is expected to be
measurable.  When the errors in parallax (and consequently in proper
motion) are increased to 10\%, only 1 of our 15 streams can be used to
determine the time evolution. Reasonable additional cuts in distance
errors and in frequency space (because this is most strongly affected
by errors), allowed the time evolution to be determined for an additional four streams. The best results were obtained for streams that are very radial and have small pericentres (inside the solar circle). This is
predominantly because these streams have many stars 
close to the Sun and therefore have smaller observational uncertainties.
There is no guarantee that streams with such characteristics 
will be present in the Gaia dataset, of course, both because of the requirement on their orbital properties and on the rather optimistic assumptions on the distance 
and proper motion errors.

Another important issue is that the correct potential is necessary to
derive the correct angle and frequency space
distributions. Fortunately, an incorrect potential distorts the
distribution in angle and in frequency spaces in such a manner that
each space can be used independently to improve the estimation of the
potential for sufficiently long streams. For short streams (e.g.\ with
only one wrap), the misalignment observed might be confused with time
evolution instead of the incorrect assumptions about the
potential. The use of short streams might therefore induce a bias in
the derived values of the characteristic parameters of the potential
or its time dependence.

By construction, we neglected the effect of self-gravity in our experiments. Self-gravity would create a gap in the action space between (the progenitor and) the leading and trailing streams \citep{Gibbons2014}, because the particles released at specific points along the orbit (rather than continuously) are offset in energy \citep{Johnston1998}. This structure will most likely also be apparent in the angle and frequency distributions of the particles that make up the stream \citep{Eyre2011}. Moreover, when close to the progenitor, released particles still experience a gravitational pull by the progenitor \citep{Choi2009}, and epicyclic oscillations are seen in streams in N-body simulations
\citep{Kuepper2010,Kuepper2012}. These effects may complicate
fitting a straight line to the distribution and determining its slope accurately, but after the particles become unbound, the dynamics is essentially the same as in our test-particle integrations \citep{Sanders2013a}.

Other interesting time-dependent effects may be involved in shaping streams and complicate interpretations. Interactions with dark subhaloes orbiting the Galactic halo will affect the 
structure of streams and create gaps, for example \citep{Yoon2011, Carlberg2013, Ngan2014}. 
Dwarf galaxies as massive as the Large and Small Magellanic Clouds are able to significantly perturb the potential \citep{VeraCiroHelmi2013}, and might induce non-adiabatic changes in the stream orbits \citep{Gomez2015}. It therefore seems important to try and understand these less secular effects on the dynamical evolution of streams.

We have presented here the first steps towards understanding the
imprints of time evolution. More explorations are necessary because we did not study other density profiles or deviations
from spherical symmetry, nor did we take into account the effects of a live halo or a (growing) disk. It may be possible to extend our method to any
potential for which action-angle coordinates can be derived or
approximated, and we do expect the general behaviour to be similar when the potential
is growing adiabatically. Overall, we expect the imprint of the smooth mass growth of the Galactic dark halo to be present in streams and to a have a small but non-negligible magnitude.\\

H.J.T.B. and A.H. gratefully acknowledge financial support from
ERC-Starting Grant GALACTICA-240271. We thank the anonymous referee for a constructive report 
that helped improve this manuscript. H.J.T.B. thanks Robyn
E. Sanderson for the many interesting and useful discussions.

\bibliographystyle{aa}
\bibliography{article}

\appendix
\section{Computing the angles}
\label{sec:AppendixA}
The angles result from the derivatives of the generating function $W(\vec{x},\vec{J})$ w.r.t.\ the actions
\begin{equation}
  \boldsymbol{\theta} = \frac{\partial W(\vec{x},\vec{J})}{\partial \vec{J}} ,
\end{equation}
\citep[Eq.~3.204]{BinneyTremaine2008}. The complete generating function for a spherical system is given by \citep[modified from Eq.~3.220, ][]{BinneyTremaine2008}
\begin{equation}
\begin{split}
  W(\vec{x},\vec{J}) &= W_\phi(\phi, \vec{J}) + W_\vartheta(\vartheta, \vec{J}) + W_r(r, \vec{J}) \\
                           &= \phi J_\phi + \int_{\vartheta_\text{min}}^\vartheta d\vartheta\, p_\vartheta(J_\phi, J_\vartheta) + \int_{r_\textrm{peri}}^r dr\, p_r(J_r, J_\phi, J_\vartheta) ,
\end{split}
\end{equation}
where we integrate over the particles trajectory in phase space, which means for example that during a whole radial period, the particle transverses twice the branch from pericentre to apocentre, but once in reversed direction. The latitudinal momentum $p_\vartheta$ is
\begin{equation}
  p_\vartheta^2 = L^2 - \frac{L_z^2}{\sin^2 \vartheta} = L^2 \left( 1 - \frac{\cos^2 i }{\sin^2 \vartheta} \right) ,
\end{equation}
where $\cos i = L_z / L$. The radial momentum $p_r$ is
\begin{equation}
  p_r^2 = 2\left(H(\vec{J}) - \Phi(r) \right) - \frac{L^2}{r^2} ,
\end{equation}
where $L = J_\theta + |J_\phi|$ and $J_\phi \geq 0$ is assumed\footnote{Otherwise, every derivative of $L$ w.r.t.\ $J_\phi$ results in a multiplication with $\textrm{sign}(J_\phi)$.}. \\

We find the radial angle as
\begin{equation}
 \theta_r = \frac{\partial W(\vec{x},\,\vec{J})}{\partial J_r} = \frac{\partial H}{\partial J_r} \frac{\partial W_r}{\partial H} = \Omega_r W_{r,\,H}, \pmod{2\pi} ,
\end{equation}
where we note that the angles are always defined modulo $2\pi$ because there is no information on how many loops a particle made around the Galactic centre. The derivative of the generating function is given by
\begin{align}
\nonumber W_{r,\,H} = \int \frac{dr}{p_r} &= \begin{cases}
   f_1(r) & \text{if } p_r \geq 0 , \\[4pt]
   2 f_1(r_\text{apo}) - f_1(r) & \text{if } p_r < 0 ,
 \end{cases}  \\[4pt]
  \nonumber f_1(r) &= \int_{r_\text{peri}}^r \frac{dr}{p_r} , \\[4pt]
  f_1(r_\text{apo}) &= \frac{\pi}{\Omega_r} ,
\end{align}
where the conditions on $p_r$ are necessary to take the right branch of $f_1$. After one full period, we find $\theta_r = 2\pi$ as expected. \\

For the azimuthal angle we find
\begin{equation}
\begin{split}
  \theta_\phi = \frac{\partial W(\vec{x},\,\vec{J})}{\partial J_\phi} = \phi + W_{\vartheta,\,J_\phi} + W_{r,\,J_\phi} + \Omega_\phi W_{r,\,H},  \pmod{2\pi} ,
\end{split}
\end{equation}
where we find $\phi$ using the (quadrant-aware) arctangent
\begin{equation}
  \phi = \arctan(y/x) .
\end{equation}
The derivative $W_{r,\,H}$ has already been worked out for $\theta_r$, and the other derivatives are
\begin{align}
        \nonumber  W_{r,\,J_\phi} = - L \int \frac{dr} {p_r \, r^2} &= - L \begin{cases} 
   f_2(r) & \text{if } p_r \geq 0 , \\[4pt]
   2 f_2(r_\text{apo}) - f_2(r) & \text{if } p_r < 0 ,
 \end{cases}\\[4pt]
           \nonumber f_2(r) &= \int_{r_\textrm{peri}}^r \frac{dr} {p_r \, r^2} , \\[4pt]
           f_2({r_\textrm{apo}}) &= \frac{\Omega_\phi}{L} f_1(r_\text{apo}) ,
\end{align}
\begin{align}
    \nonumber W_{\vartheta,\,J_\phi} = \int \frac{d\vartheta}{p_\vartheta} \left[ L - \frac{J_\phi}{\sin^2\vartheta} \right] &= \int d\vartheta \frac{ 1 - \frac{\cos i}{\sin^2\vartheta}}{\sqrt{1 - \frac{\cos^2 i }{\sin^2 \vartheta}}}
    = \begin{cases}
   f_3(\vartheta) & \text{if } p_\vartheta \geq 0 , \\[4pt]
   -f_3(\vartheta) & \text{if } p_\vartheta < 0 , 
 \end{cases} \\[4pt]
         \nonumber f_3(\vartheta) &= \int_{\vartheta_\text{min}}^\vartheta  d\vartheta \frac{ 1 - \frac{\cos i}{\sin^2\vartheta}}{\sqrt{1 - \frac{\cos^2 i }{\sin^2 \vartheta}}} , \\[4pt]
         f_3(\vartheta_\text{max}) &= 0.
\end{align}
The function $W_{\vartheta,\,J_\phi}$ is oscillatory in nature (w.r.t.\ $\vartheta$), while the combination $W_{r,\,J_\phi} + \Omega_\phi W_{r,\,H}$ is also oscillatory (w.r.t.\ $r$): after one radial period it evaluates to $-2Lf_2(r_\text{apo}) +2 \Omega_\phi f_1(r_\text{apo}) = 0$.

The latitudinal angle is given by
\begin{equation}
\begin{split}
  \theta_\vartheta = \frac{\partial W(\vec{x},\,\vec{J})}{\partial J_\vartheta} = W_{\vartheta,\,J_\vartheta} + W_{r,\,J_\vartheta} + \Omega_\vartheta W_{r,\,H}, \pmod{2\pi} .
\end{split}
\end{equation}
Because we assume $J_\phi \geq 0$, we find $W_{r,\,J_\vartheta} = W_{r,\,J_\phi}$ and $\Omega_\vartheta = \Omega_\phi$. The remaining derivative of the generating function is
\begin{align}
    \nonumber W_{\vartheta,\,J_\vartheta} = \int \frac{d\vartheta}{p_\vartheta} L &= \int d\vartheta \frac{ 1 }{\sqrt{1 - \frac{\cos^2 i }{\sin^2 \vartheta}}} = \begin{cases}
   f_4(\vartheta) & \text{if } p_\vartheta \geq 0 , \\[4pt]
   2\pi-f_4(\vartheta) & \text{if } p_\vartheta < 0 ,
 \end{cases} \\[4pt]
         \nonumber f_4(\vartheta) &= \int_{\vartheta_\text{min}}^\vartheta d\vartheta \frac{ 1 }{\sqrt{1 - \frac{\cos^2 i }{\sin^2 \vartheta}}} ,\\[4pt]
         f_4(\vartheta_\text{max}) &= \pi ,
\end{align}
where the term $W_{r,\,J_\vartheta} + \Omega_\vartheta W_{r,\,H}$ vanishes after one radial period, while the term $W_{\vartheta,\,J_\vartheta}$ contains the dependence on $\vartheta$ and increases by $2\pi$ after one period in $\vartheta$.

\section{Transformation equations}
\label{sec:AppendixB}
The linearised transformation between action-angle coordinates and Cartesian coordinates is
\begin{equation}
	\vec{T} = \left[ \renewcommand\arraystretch{2.2}\begin{matrix} \dfrac{\partial^2 W}{\partial \vec{J} \partial \vec{J}} \dfrac{\partial \vec{J}}{\partial \vec{q}} + \dfrac{\partial^2 W}{\partial \vec{J} \partial \vec{q}} & \dfrac{\partial^2 W}{\partial \vec{J} \partial \vec{J}} \dfrac{\partial \vec{J}}{\partial \vec{p}} \\ \dfrac{\partial \vec{J}}{\partial \vec{q}} & \dfrac{\partial \vec{J}}{\partial \vec{p}}
\end{matrix} \right] .
\end{equation}
For simplicity, we provide here the matrix for the 2D case (i.e.\ when the orbit is in the plane). In that case, $J_\theta = 0$ and $J_\phi = L = L_z$, so that we find
\begin{equation}
	\vec{T} = \left[ \renewcommand\arraystretch{2.2}\begin{matrix} 				
	1
	& t_{\theta_\phi,\,r}
	& t_{\theta_\phi,\,p_\phi}
	& t_{\theta_\phi,\,p_r} \\
	0 
	& t_{\theta_r,\,r}
	& t_{\theta_r,\,p_\phi}
	& t_{\theta_r,\,p_r} \\
	0
	& 0
	& 1
	& 0 \\
	0
	& t_{J_r,\,r}
	& t_{J_r,\,p_\phi}
	& t_{J_r,\,p_r}
\end{matrix} \right] ,
\end{equation}
where the above terms are given by
\begin{align}
	& t_{\theta_\phi,\,r} = \frac{\kappa}{p_r}  + W_{J_\phi,\,J_r} \frac{\eta}{\Omega_r} ,
	& t_{\theta_\phi,\,p_\phi} &= W_{J_\phi,\,J_\phi} - W_{J_\phi,\,J_r}\frac{\kappa}{\Omega_r} , \\
	& t_{\theta_\phi,\,p_r} = W_{J_\phi,\,J_r}  \frac{p_r}{\Omega_r} , 
	& t_{\theta_r,\,r} &= \frac{\Omega_r}{p_r} + W_{J_\phi,\,J_\phi} \frac{\eta}{\Omega_r} , \\
	& t_{\theta_r,\,p_\phi} = W_{J_\phi,\,J_r} - W_{J_r,\,J_r} \frac{\kappa}{\Omega_r} , 
	& t_{\theta_r,\,p_r} &= W_{J_r,\,J_r}  \frac{p_r}{\Omega_r} ,  \\
	& t_{J_r,\,r}    = \frac{\eta}{\Omega_r} , 
	& t_{J_r,\,p_\phi} &= -\frac{\kappa}{\Omega_r} , \\
	& t_{J_r,\,p_r} = \frac{p_r}{\Omega_r} . 
\end{align}
The functions $\kappa$ and $\eta$ are
\begin{align}	
	\kappa &= \Omega_\phi - \frac{L}{r^2} ,\\
	\eta &= \diffp{\Phi}{r} - \frac{L^2}{r^3} ,
\end{align}
and the $W_{J_i,\,J_j}$ are found by differentiating the generating function
\begin{align}
	W_{J_\phi,\,J_\phi} &= \diffp{W}{{J_\phi^2}} =  \int_{r_\textrm{peri}}^r \frac{dr}{p_r} \left( \diffp{\Omega_\phi}{{J_\phi}} - \frac{1}{r^2} - \frac{\kappa^2}{p_r^2} \right), \\
	W_{J_r,\,J_r} &= \diffp{W}{{J_r}{J_r}} =  \int_{r_\textrm{peri}}^r \frac{dr}{p_r} \left( \diffp{\Omega_r}{{J_r}} - \frac{\Omega_r^2}{p_r^2} \right), \\
	W_{J_r,\,J_\phi} &= \diffp{W}{{J_r}{J_\phi}} =  \int_{r_\textrm{peri}}^r \frac{dr}{p_r} \left( \diffp{\Omega_\phi}{{J_r}} - \frac{\kappa}{p_r^2}\Omega_r \right).
\end{align}

\end{document}